\documentclass[conference]{IEEEtran}
\IEEEoverridecommandlockouts
\usepackage{cite}
\usepackage{amsmath,amssymb,amsfonts}
\usepackage{graphicx}
\usepackage{textcomp}
\usepackage{xcolor}
\def\BibTeX{{\rm B\kern-.05em{\sc i\kern-.025em b}\kern-.08em
    T\kern-.1667em\lower.7ex\hbox{E}\kern-.125emX}}
    \usepackage{xcolor}
\usepackage{amsmath}
\usepackage{graphicx}
\usepackage{float}
\usepackage{comment}
\usepackage{algorithmic}
\usepackage[linesnumbered,ruled]{algorithm2e}
\usepackage{titlesec}
\usepackage{multirow}
\usepackage{array}
\usepackage{tabularx}
\usepackage{optidef} % For optimization problem formatting
 
\usepackage{booktabs}
\usepackage{amssymb}
\usepackage{booktabs}
\usepackage{graphicx}
\usepackage{subcaption}
\usepackage{float}
\usepackage{url}

\DeclarePairedDelimiter\floor{\lfloor}{\rfloor}% For coloring the DNC rows

\begin{document}

\title{AC-aware Optimization Framework for Under-Frequency Load Shedding 
}
\author{
\IEEEauthorblockN{Mazen Elsaadany, Muhammad Hamza Ali, Amritanshu Pandey, 
Mads R.Almassalkhi}
\IEEEauthorblockA{Department of Electrical and Biomedical Engineering \\
University of Vermont\\
Burlington, Vermont}
}
% \and
% \and
% \and
% \and
% \and

\maketitle

\begin{abstract}

Under-frequency load shedding (UFLS) prevents system collapse during large disturbances. Increased penetration of distributed energy resources (DERs) and reduced system inertia makes it challenging to design a static UFLS scheme, which relies on preset frequency thresholds and load shed fractions to meet design criteria across all possible operating conditions. Due to non-linearity and traceability issues, previous adaptive UFLS schemes use simplified tractable frequency models that overlook AC network effects such as voltage-dependent load/generation. This paper leverages model order reduction techniques to obtain a higher fidelity low-order model of system frequency dynamics that captures AC network effects while incorporating turbine governor action and their associated limits. The model is then used in a new AC-aware predictive optimization framework to adapt UFLS setpoints periodically based on current operating conditions while minimizing load shed. Validated on a 1,648-bus system with PSS/E simulations, the proposed method meets design criteria under various operating conditions and disturbance scenarios. Furthermore, the framework outperforms conventional static UFLS schemes and adaptive UFLS schemes based on simplified dynamic models.

\end{abstract}

\begin{IEEEkeywords}
    Power system dynamics, Model Order Reduction, Under-frequency Load Shedding, Mixed Integer Program.
\end{IEEEkeywords}

\section{INTRODUCTION} \label{sec:Introduction}
The transition of the power grid towards renewable energy resources (RES) has significantly reshaped the operation and dynamics of modern power systems. The integration of RES technologies, such as wind and solar photovoltaic systems (PV), has displaced traditional synchronous generators, leading to a reduction in system inertia traditionally provided by synchronous generators \cite{kabeyi2022sustainable}. Reduced inertia levels lead to larger frequency sensitivity to power imbalances, impacting grid reliability and stability. 
Under-frequency load shedding (UFLS) is used as the power grid's emergency brake and sheds load during large contingencies to help restore the balance between load and generation, halt frequency decline, and help prevent overall system collapse. Conventional UFLS schemes are static and rely on several UFLS relays in the network with preset frequency thresholds. The UFLS relays are designed such that when they trip a predetermined percentage of load is shed when the frequency at the relay drops below a preset threshold \cite{alhelou2020overview}. Typically, UFLS schemes consist of multiple stages each with a corresponding frequency threshold and load shed amount. The amount of load shed and the frequency thresholds are designed to meet certain performance criteria laid out by the North American Electric Reliability Corporation (NERC) for contingencies with an imbalance of up to 25\% between load and generation\cite{NERC}.
However, with the increased prevalence of distributed energy resources (DERs), load patterns behind the UFLS relays have become more erratic with larger variations\cite{load_availability_for_UFLS_Australia}. Furthermore, generation from DERs such as residential solar PV can exceed local demand leading to reversed power flow (backfeeding). Static UFLS schemes do not adapt to changing operating conditions, fluctuating load patterns, reduced system inertia, and bi-directional power flows. As a result, static UFLS schemes may fail to restore frequency to within safe limits.

% , operating conditions may be far away from the design conditions under which the UFLS scheme was designed
% This bidirectional power flow coupled with increasingly more fluctuating load patterns and reduced system inertia makes it challenging to design a static UFLS scheme that can reliably meet design criteria under all possible operating conditions\cite{UFLS_overview_paper}. 
% Furthermore, with the increasing penetration of DERs, and the reduced inertia caused by RESs displacing traditional synchronous generators, operating conditions may be far away from the design conditions under which the UFLS scheme was designed. This can lead to over-shedding or under-shedding of load, leading to either excessive consumer power interruptions or insufficient frequency recovery which may cause cascading failures and even blackouts.

Therefore, there is a need for an adaptive UFLS scheme that can respond to changing grid conditions. Previous work on adaptive UFLS schemes can be broadly classified as either response-based or prediction-based. In response-based schemes \cite{response_based_method,response_method_2,WAMS_based_UFLS} the magnitude of the disturbance/contingency is estimated via the overall system rate of change of frequency (RoCoF) and is used to determine the amount of load to be shed. Response-based schemes use the center of inertia (CoI) frequency/RoCoF along with network and generator parameters values to estimate the magnitude of the contingency and shed load accordingly \cite{response_method_2,response_based_method}. This requires measurements from all generators in the network which begets a centralized high-speed reliable communication infrastructure to estimate RoCoF in real-time \cite{WAMS_based_UFLS}. Distributed approaches to estimate system RoCoF from local measurements have been proposed \cite{Rocof_estimation,Rocof_estimation_algorithms_for_UFLS}. However, response-based methods are prone to error due to noise and large frequency swings associated with low inertia systems and high penetration of inverter interfaced renewable energy resources\cite{UFLS_overview_paper}.

In prediction-based schemes \cite{MILP_UFLS_2,MILP_UFLS_Probabilistic,FeiTeng_SFR_based_constraints}, a predictive model of the system's frequency dynamics is used to optimize UFLS setpoints. Prediction-based schemes commonly use the simplified frequency response model (SFR), presented in \cite{SFR_Original,SFR_addition}, as a proxy for the system's frequency dynamics. This is done to facilitate analysis and/or arrive at a tractable UFLS optimization formulation\cite{MILP_UFLS_2,MILP_UFLS_Probabilistic,FeiTeng_SFR_based_constraints}. However, the SFR model is a simplified aggregated model of the system's frequency dynamics that represents the system with a single equivalent generator and in doing so ignores the AC network and the effects of voltages on frequency dynamics made more evident with the presence of voltage-dependent loads. Previous work that does consider network information has been restricted to DC power flow based models and for continuously controllable resources such as inverters providing virtual inertia and/or damping \cite{Poolla_Dorfler_virtual_Inertia}. UFLS, on the other hand, consists of discrete load-shedding control actions.
Work presented in \cite{MILP_UFLS_Probabilistic,MILP_UFLS_2} use the SFR model to formulate a Mixed Integer Linear Program (MILP) that optimizes UFLS setpoints while ensuring that the time domain frequency response of the system meets certain design criteria. Work presented in \cite{FeiTeng_SFR_based_constraints} presents a comprehensive optimization framework that co-optimizes UFLS setpoints along with slow and fast frequency regulation services and system inertia. The work in~\cite{FeiTeng_SFR_based_constraints} does not use a time-domain approach but uses the SFR model to formulate explicit constraints on system frequency nadir and uses that to optimize UFLS setpoints along with other frequency services. Furthermore, previous work has largely omitted turbine governor limits which greatly impact frequency response \cite{sauer_power_nodate}, especially during large contingencies which is the case in  UFLS applications.

Despite not suffering from the same practical implementation challenges as response-based methods, prediction-based methods have mainly utilized the SFR model as the predictive model of the system's frequency dynamics and largely ignored governor limits. Hence, this paper presents a prediction-based UFLS optimization scheme that adapts slow coherency-based reduction methods to arrive at a low-order frequency model of system frequency dynamics that is AC-aware (captures the dependency of frequency dynamics on bus voltages)\cite{Joe_Chow_MOR_book,Dorfler_slow_coherency,Slow_Coherency}. The prediction-based UFLS scheme takes in information about the network, generators, governor limits, and grid operating conditions to optimize UFLS setpoints for the largest credible contingency (25\% imbalance) such that to minimize the amount of load shed and ensure satisfactory performance criteria as laid out by NERC \cite{NERC}. Additionally, constraints are added to prevent excessive load shedding at any given UFLS stage. This is to ensure no over shedding in case of smaller disturbances.

Th main contributions of the paper are as follows:
\begin{enumerate}
    \item \textbf{Higher model fidelity}: We leverage techniques used in slow-coherency-based model order reduction to obtain a Simplified AC-aware Frequency Response (SAFR) model of system frequency dynamics. Due to the higher model fidelity compared to the SFR model, predictive-based optimization of UFLS setpoints using the SAFR model outperforms SFR-based UFLS optimization schemes as well as static UFLS schemes.

    % The UFLS optimization scheme utilizes a MOR-based predictive model that includes the effect of voltage-dependent loads and generator power injections, unlike the SFR model. Furthermore, the formulation captures the effects of governor saturation. The ability of the proposed UFLS scheme to adapt to changing conditions, coupled with the higher fidelity modeling, causes it to outperform static UFLS schemes as well as SFR-based UFLS schemes.
    \item \textbf{Scalability to Larger Systems}: The proposed SAFR model is a low-order dynamic model conducive to scaling the proposed prediction-based UFLS optimization for larger networks.
    \item \textbf{Empirical Validation}: The performance of the optimization framework is validated against full non-linear time-domain simulations run on PSS/E on a 1648 bus system. The optimization framework is shown to meet design criteria under various operating conditions and disturbance scenarios. Furthermore, the proposed scheme outperforms static UFLS schemes (state-of-practice) as well as SFR-based UFLS schemes.
\end{enumerate}

The rest of the paper is organized as follows: Section~\ref{sec:System_modeling} presents the full-order system model and the SFR model and highlights the pitfalls of using the SFR model. Section~\ref{sec:MOR} presents the slow-coherency aggregation techniques used to arrive at a low-order model of the system's frequency dynamics (SAFR) and compares the low-order SAFR model to the full-order non-linear model. Section~\ref{sec:UFLS Setpoint Optimization} presents the MILP formulation of the UFLS setpoint optimization framework. Section~\ref{sec:Simulation_Res} presents the simulation results and validation of the optimization framework. Finally, Section~\ref{sec:Conclusion} concludes the paper.

\section{Overview of System Modeling}\label{sec:System_modeling}
We consider a power system electromechanical model to capture generator frequency dynamics and bus voltage behavior. The model is then used to predict the frequency response to optimize UFLS setpoints. Typically, the tractable low-order SFR model is used~\cite{FeiTeng_SFR_based_constraints,MILP_UFLS_2,MILP_UFLS_Probabilistic}. The SFR model, however, ignores AC effects  (bus voltages, line flows, etc.) on the frequency dynamics. To remedy this, we adapt techniques from slow coherency model order reduction to obtain the SAFR model from the linearized full order system. A full-order power system dynamics model is first presented and then the SFR model is given and the response of both models and the linearized dynamics to a loss of generation contingency are compared. 

\subsection{Full-order Power System Dynamics Model}
Consider a power system comprised of $N_\text{g}$ generators and $N$ buses. Let $\mathcal{G}:=\{1,\hdots, N_\text{g}\}$ denote the set of generators with rotor angle and frequency are denoted by $\delta, \omega \in \mathbb{R}^{N_\text{g}}$ and let $\mathcal{B}:=\{1,\hdots,N\}$ denote the set of all buses with the bus voltage magnitudes and phase angles denoted as $V, \theta \in \mathbb{R}^N$. %Let $\delta_\text{g}$, $\omega_\text{g} \in \mathbb{R}^{N_\text{g}}$ be vectors that contain the individual generator rotor angle and frequencies $\delta_{\text{g},i}$ and $\omega_{\text{g},i}$ respectively. 
%Similarly, let $\delta_\text{b}$, $V_\text{b} \in \mathbb{R}^N$ be vectors that contain the individual bus voltage angles and magnitudes $V_{\text{b},n}$ and $\delta_{\text{b},n}$ respectively. 

% The variables $ \left(\delta_\text{g},\quad\omega_\text{g}\right)$ are governed by the generator swing equations coupled to the network via generator injections, and the power flow equations dictate the algebraic variables $ \left(\delta_\text{b}\quad V_\text{b}\right)$.  

% Note that the deviations in $\left(\delta_\text{g},\omega_\text{g},\delta_{\text{b}},V_{\text{b}} \right)$ from nominal is considered, however, the same notation is used to simplify notation.
%%%%%%%%%%%%%%%%%%%%%%%%%%%%%%%%%%%%%%%%%%%%%%%%%%%%%%%%%%%%%%%%%%%%%%%%%%%%%%%%%%%%%%%%%%%%%%%%%%%%%%%%%%%%%%%%%%%%%%%%%%%%%%%%%%%%%%%%%%%%%%%%%%%%%%%%%%%%%%%%%%%%%%%%
 \subsubsection{Generator Swing Dynamics}
 The classical electromechanical model \cite{sauer_power_nodate,kundur_power_1994} is considered wherein the generator $i$ is modeled as a constant internal voltage $E_i$ behind a transient reactance $X^{\prime}_{\text{d},i}$. The dynamics of machine $i\in \mathcal{G}$ located at bus $j \in \mathcal{B}$ is modeled as
\begin{subequations}\label{eqn:swing_eqn}
 \begin{align}
 \dot{\delta}_{i} &= {\omega}_{i}\\
     \frac{m_i}{\omega_0}\dot{\omega}_{i} &= -d_i{\omega}_{i}+P_{\text{m},i} -P_{\text{e},i} \\ P_{\text{e},i}&=\frac{V_{j}E_i\sin(\delta_{i}-\theta_{j})}{X^{\prime}_{\text{d},i}} \\
      Q_{\text{e},i} &= \frac{V_{j}E_i}{X^{\prime}_{\text{d},i}}\cos\left(\delta_{i} - \theta_{j} \right) - \frac{V_{j}^2}{X^{\prime}_{\text{d},i}},
 \end{align}
\end{subequations}
\noindent where $m_i$ is the generator inertia constant, $d_i$ is the generator damping coefficient,$\omega_0$ is the nominal frequency in rad/s, $P_{\text{m},i}$ is the mechanical power input to the generator, and $P_{\text{e},i}$ and $Q_{\text{e},i}$ are the real and reactive output power to the grid, and $V_{j}$ and $\theta_{j}$ are the terminal bus voltage magnitude and phase angle of bus $j$. 

% Furthermore, the rotor angle is measured with respect to a stationary reference frame in \eqref{eqn:swing_eqn}. A rotating reference frame can also be used, this would transform the rotor angles as shown below:
% \begin{subequations}
%     \begin{align}
%         {\delta}_{\text{g},i}^\prime &= {\delta}_{\text{g},i} - {\delta}_\text{ref}\\
%         \dot{\delta}_{\text{g},i}^\prime &= {\omega}_{\text{g},i} - {\omega}_{\text{ref}},
%     \end{align}
% \end{subequations}
% \noindent where ${\delta}_\text{ref}$ and ${\omega}_{\text{ref}}$ are the reference angle and frequency.

%%%%%%%%%%%%%%%%%%%%%%%%%%%%%%%%%%%%%%%%%%%%%%%%%%%%%%%%%%%%%%%%%%%%%%%%%%%%%%%%%%%%%%%%%%%%%%%%%%%%%%%%%%%%%%%%%%%%%%%%%%%%%%%%%%%%%%%%%%%%%%%%%%%%%%%%%%%%%%%%%%%%%%%%

\subsubsection{Turbine-Governor Dynamics}
The mechanical power input $P_{\text{m},i}$ to machine $i \in \mathcal{G}$ is determined by the turbine governor which is modeled using a first-order low-pass filter~\cite{Poola_Dorfler_VI_book} as
\begin{align}\label{eqn:Governor_TF}
 P_{\text{m},i}(s)  = -\frac{{\omega}_{i} - \omega_0}{R_i(T_{i} s + 1)} + P_{\text{m},i}^\text{ref} \qquad \forall i\in \mathcal{G},
\end{align}
\noindent where $\omega_0$ is the nominal frequency, $P_{\text{m},i}^\text{ref}$ is the the nominal value of mechanical power input, $R_i$ is the governor droop gain and $T_{i}$ is the governor time constant. From~\eqref{eqn:Governor_TF} the following is obtained:
\begin{align}\label{eqn:Governor ODE}
\dot{P}_{\text{m},i}=\frac{1}{T_{i}}\left (-\frac{\omega_{i}-\omega_0}{R_i} - {P}_{\text{m},i} +P_{\text{m},i}^\text{ref}\right) \qquad \forall i\in \mathcal{G} .
\end{align}

%%%%%%%%%%%%%%%%%%%%%%%%%%%%%%%%%%%%%%%%%%%%%%%%%%%%%%%%%%%%%%%%%%%%%%%%%%%%%%%%%%%%%%%%%%%%%%%%%%%%%%%%%%%%%%%%%%%%%%%%%%%%%%%%%%%%%%%%%%%%%%%%%%%%%%%%%%%%%%%%%%%%%%%%

 \subsubsection{Transmission Network}
 The voltage phasors $V_{n}\angle \theta_{n}$ satisfy power balance equations $\forall n\in \mathcal{B}$, which is summarized with the AC power flow:
 \begin{subequations}\label{eqn:pf_eqns}
\begin{align}
P_{\text{net},n} &= V_n \sum_{k=1}^{N} V_k \left[ G_{nk} \cos(\theta_{nk}) + B_{nk} \sin(\theta_{nk}) \right] \label{eqn:pf_eqns_P} \\
  Q_{\text{net},n} &= V_n \sum_{k=1}^{N} V_k \left[ G_{nk} \sin(\theta_{nk}) - B_{nk} \cos(\theta_{nk}) \right]\label{eqn:pf_eqns_Q},
% P_n &= V_n \sum_{k=1}^{N} V_k \left[ G_{nk} \cos(\delta_{\text{b},n} - \delta_{\text{b},k}) + B_{nk} \sin(\delta_{\text{b},n} - \delta_{\text{b},k}) \right] \label{eqn:pf_eqns_P} \\
%   Q_n &= V_n \sum_{k=1}^{N} V_k \left[ G_{nk} \sin(\delta_{\text{b},n} - \delta_{\text{b},k}) - B_{nk} \cos(\delta_{\text{b},n} - \delta_{\text{b},k}) \right]\label{eqn:pf_eqns_Q},
\end{align}
 \end{subequations}
where $G_{nk}$ and $B_{nk}$ are the conductance and susceptance for line $(n,k)$ while $\theta_{nk} := \theta_{n}-\theta_{k}$ is the phase angle difference between buses $n$ and $k$. The left hand sides of~\eqref{eqn:pf_eqns} represent the net power injection at bus $n$ and are given by,
\begin{subequations}\label{eqn:pf_net_injections}
    \begin{align}
        P_{\text{net},n} &:= P_{\text{g},n} - P_{\text{d},n}\\
        Q_{\text{net},n} &:= Q_{\text{g},n} - Q_{\text{d},n},
    \end{align}
\end{subequations}
 where $P_{\text{g/d},n}$ and $Q_{\text{g/d},n}$ are active and reactive power injections/demand at buses $n$, respectively. %Similarly,  $P_{\text{d},n}$ and $Q_{\text{d},n}$ are the real and reactive power demand at bus $n$. Generator power injections are not constant but functions of  $\delta_{\text{g}}, \omega_{\text{g}},\delta_{\text{b}}$, and $V_{\text{b}}$. The generator real and reactive power injections at bus $n$ are given by , 
%  \begin{subequations}\label{eqn:Generator PQ injections}
%     \begin{align}
%         P_{\text{g},n} = \frac{V_{\text{b},n}E_i}{X^{\prime}_{\text{d},i}}\sin\left(\delta_{\text{g},i} - \delta_{\text{b},n} \right) \\
%         Q_n = \frac{V_{\text{b},n}E_i}{X^{\prime}_{\text{d},i}}\cos\left(\delta_{\text{g},i} - \delta_{\text{b},n} \right) - \frac{V_{\text{b},n}^2}{X^{\prime}_{\text{d},i}},
%     \end{align}
% \end{subequations}
% where machine $i$ is connected to bus $n$.
The power demand $\left(P_{\text{d},n},Q_{\text{d},n}\right)$ are modeled as voltage-dependent ZIP loads with constant power, current, and admittance components.   $P_{\text{d},n}$ and $Q_{\text{d},n}$ are given as follows: %functions of $V_{\text{b}}$
\begin{subequations}\label{eqn:ZIP_load}
    \begin{align}
        P_{\text{d},n} &= P_{\text{P},n} + V_{n}I_{\text{P},n} + V^2_{n}Y_{\text{P},n}\\
        Q_{\text{d},n} &= Q_{\text{Q},n} + V_{n}I_{\text{Q},n} + V^2_{n}Y_{\text{Q},n},
    \end{align}
\end{subequations}
where $P_{\text{P},n}$ and $Q_{\text{Q},n}$ are the constant active and reactive power components of the load respectively. Similarly, $I_{\text{P},n}$ and $I_{\text{Q},n}$ are the constant current components of the load and $Y_{\text{P},n}$ and $Y_{\text{Q},n}$ are the constant admittance components of the load. Given the proportions of the constant power, current, and admittance components of the ZIP load the values of $P_{\text{P},n}$, $I_{\text{P},n}$ and $Y_{\text{P},n}$ can be determined as
\begin{subequations}\label{eqn:ZIP_load_paramsP}
    \begin{align}
        P_{\text{P},n} &= (1-a_n-b_n) P_{\text{d,0},n}\\
        I_{\text{P},n} &= a_n \frac{P_{\text{d,0},n}}{V_{0,n}}\\
        Y_{\text{P},n} &= b_n \frac{P_{\text{d,0},n}}{V_{0,n}^2},
    \end{align}
\end{subequations}
where $P_{\text{d,0},n}$ and $V_{0,n}$  are the nominal power demand and voltage magnitude at bus $n$ and can be obtained from the power flow solution. Furthermore, $a_n$ and $b_n$ are the constant current and admittance fractions of the ZIP load respectively. The reactive power components $Q_{\text{Q},n}$, $I_{\text{Q},n}$ and $Y_{\text{Q},n}$ can be determined in a similar manner.

% \begin{subequations}\label{eqn:ZIP_load_paramsP}
%     \begin{align}
%         Q_{\text{Q},n} &= (1-a_n-b_n) P_{\text{d,0},n}\\
%         I_{\text{Q},n} &= a_n \frac{Q_{\text{d,0},n}}{V_{\text{b,0},n}}\\
%         Y_{\text{Q},n} &= b_n \frac{Q_{\text{d,0},n}}{V_{\text{b,0},n}^2}.
%     \end{align}
%     \end{subequations}
%%%%%%%%%%%%%%%%%%%%%%%%%%%%%%%%%%%%%%%%%%%%%%%%%%%%%%%%%%%%%%%%%%%%%%%%%%%%%%%%%%%%%%%%%%%%%%%%%%%%%%%%%%%%%%%%%%%%%%%%%%%%%%%%%%%%%%%%%%%%%%%%%%%%%%%%%%%%%%%%%%%%%%%%
\subsubsection{Full-order overall System Model}
The overall system is described by a set of differential-algebraic equations (DAE), with the dynamic variables, $x:=\left[\delta\quad\omega \quad P_{\text{m}} \right]^{T}\in \mathbb{R}^{3N_\text{g}}$, being governed by a set of ODEs~\eqref{eqn:swing_eqn} and~\eqref{eqn:Governor ODE} and the algebraic variables, $y:=\left[\theta\quad V\right]^{T}\in \mathbb{R}^{2N}$, being dictated by the power flow equations~\eqref{eqn:pf_eqns}. The overall system of DAEs can be expressed (compactly) as 
\begin{subequations}\label{eqn:DAE_compact}
    \begin{align}
        \dot{x} &= f\left(x,y\right)\\
        0 &= g\left(x,y\right),
    \end{align}
\end{subequations}
where the net power injections, $P_{\text{net},n}$ and $Q_{\text{net},n}$ are included in $g\left(x,y\right)$.
% where $I_\text{net} := \left[P_\text{net}\quad Q_\text{net}\right]^T\in \mathbb{R}^{2N}$ denotes the net active and reactive power injections at each bus.}
%%%%%%%%%%%%%%%%%%%%%%%%%%%%%%%%%%%%%%%%%%%%%%%%%%%%%%%%%%%%%%%%%%%%%%%%%%%%%%%%%%%%%%%%%%%%%%%%%%%%%%%%%%%%%%%%%%%%%%%%%%%%%%%%%%%%%%%%%%%%%%%%%%%%%%%%%%%%%%%%%%%%%%%%
\subsection{Simplified Frequency Response (SFR) Model}
\begin{figure}[t] 
    \centering
    \includegraphics[width=0.33\textwidth]{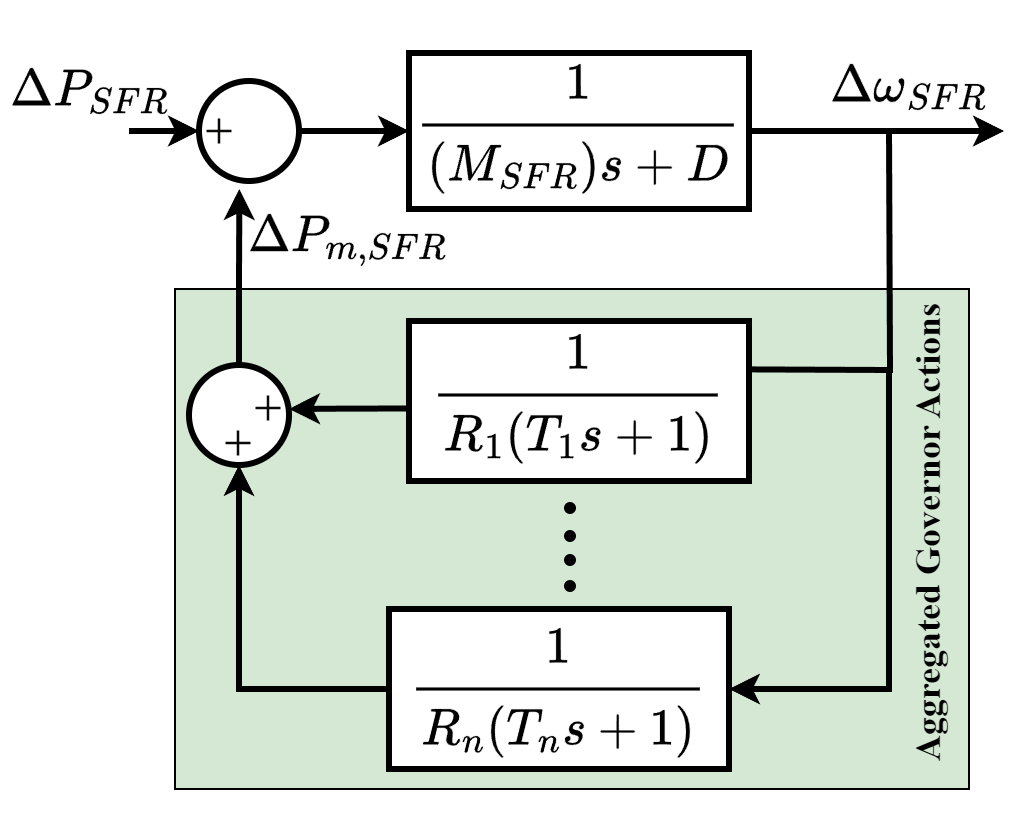}
    \caption{Simplified Frequency Response (SFR) block diagram.}
    \label{fig:SFR Block Diagram}
\end{figure}

\begin{figure}[]
    \centering
        \includegraphics[width=0.9\columnwidth]{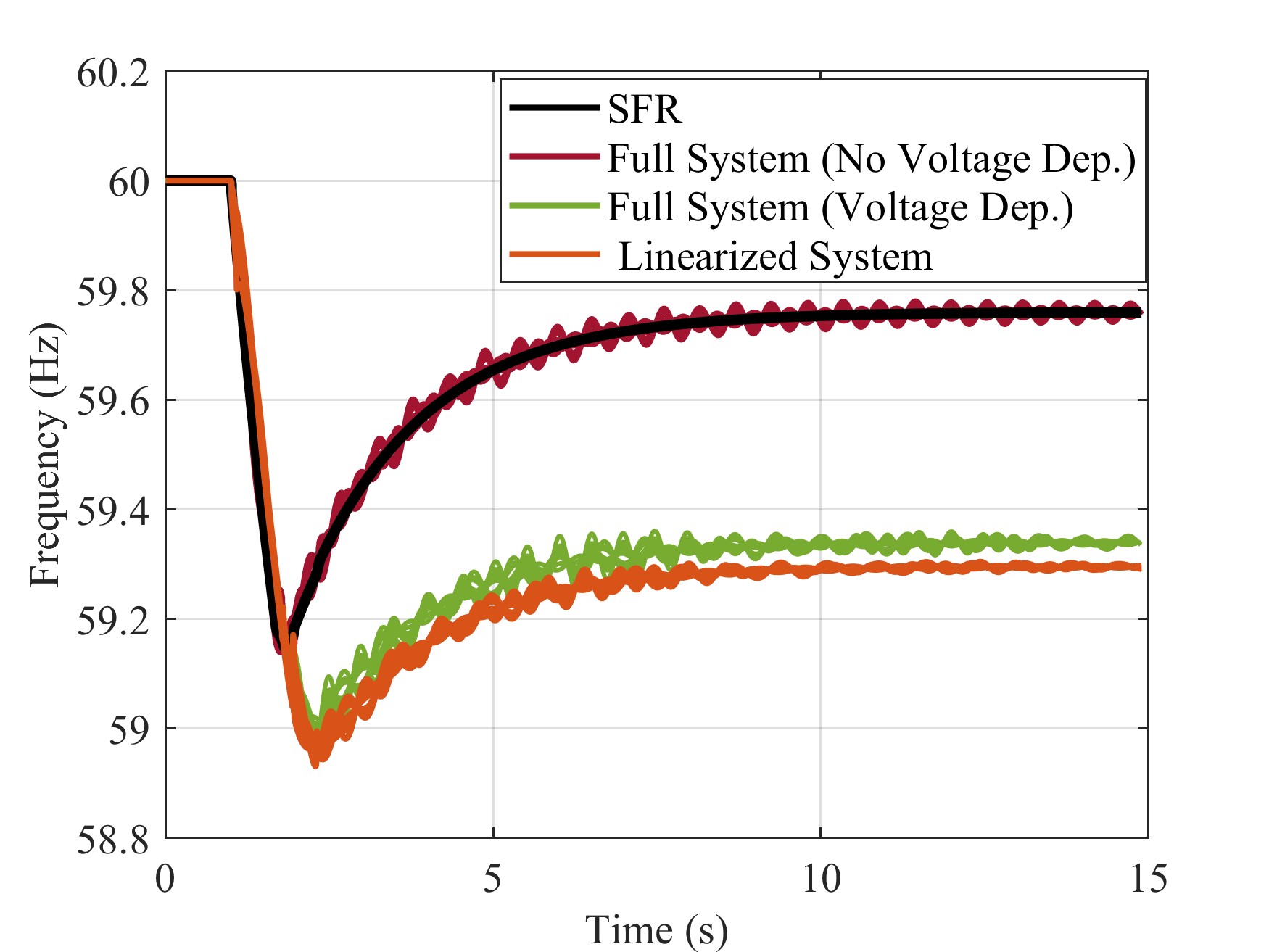}
    \caption{Comparison of frequency responses: SFR model, system response with voltage dependency, and system response without voltage dependency.}
    \label{fig:COI Comparison}
\end{figure}

The SFR model is a simplified low-order power system dynamics model that captures the averaged frequency dynamics of the system. The model is derived by averaging the individual machine dynamics in the system to obtain a single equivalent machine\cite{SFR_Original,SFR_addition}. The SFR model is illustrated in the block diagram in Fig \ref{fig:SFR Block Diagram}. $M_\text{SFR}$ and $D_\text{SFR}$ are the aggregated weighted system inertia and damping coefficient and $\Delta P_{\text{m,SFR}}$ is the total change in mechanical power input to the system, obtained by aggregating the governors as shown in Fig~\ref{fig:SFR Block Diagram}. $\Delta P_{\text{SFR}}$ is the change in total electrical power and represents the total power imbalance caused by contingencies. Note that the quantities shown in Fig~\ref{fig:SFR Block Diagram} are the changes from their respective nominal values. The SFR system frequency dynamics can be described as follows,
\begin{align}\label{eqn:SFR}
    \Delta\dot{\omega}_{\text{SFR}} = \frac{1}{M_{\text{SFR}}}\left(  -D_\text{SFR} \Delta \omega_\text{SFR} +\Delta P_{\text{m,SFR}} - \Delta P_{\text{SFR}} \right).
\end{align}
The SFR model is network agnostic and as such ignores voltage effects on load and the generator power injections, both of which affect the frequency. Figure \ref{fig:COI Comparison} considers a 23 bus test-case provided by PSS/E and shows the frequency response of the SFR model following a loss of generator contingency and UFLS action. The SFR model is compared against a) a full-order power system dynamics model based on DC power flow that does not capture voltage dependencies (as presented in \cite{Poola_Dorfler_VI_book}) , b) the non-linear model in~\eqref{eqn:DAE_compact} simulated using PSS/E , and c) a linearized version of full-order model in~\eqref{eqn:DAE_compact}. 
% shows a comparison of the bus frequency obtained from the non-linear model and the SFR model. The simulation is repeated twice, once when considering the voltage dependency of load and generator power injections and once where the voltage dependency is neglected. 
Figure~\ref{fig:COI Comparison} shows that when neglecting AC effects such voltage dependencies, the SFR model is a good approximation of full system dynamics. However, when considering the AC effects, it is evident that the SFR model fails to capture the frequency dynamics. On the other hand, the linearized model, although with some error, better captures the frequency dynamics since when considering voltage dependencies. The linearized model along with techniques borrowed from slow coherency-based aggregation methods are used to obtain the SAFR model.

% To remedy this, we next employ model order reduction techniques to obtain a reduced-order power system dynamic model that can better capture the voltage dependence of frequency dynamics.

% Typically, a simplified system frequency response model is used, originally presented in \cite{COI_OG} and later generalized to capture aggregated governor action in \cite{COI_2.0}.\mazen{This is the COI model. I'm using the wording used in the original paper}
% The simplified frequency response model averages individual machine dynamics in single equivalent machine. As such, the network is ignored and the effect of changes in $V_{\text{b},n}$ and $\delta_{\text{b},n}$  on the generator power injections and load power demand are not considered. Fig \ref{fig:COI} shows a bus frequency obtained from the non-linear model, linearized model and the simplified frequency response model. The simulation is repeated twice, once when considering load and generator power injections to be constant and once 
% when considering the voltage dependency of the load and generator power injections. 

\section{Simplified AC-aware Frequency Response Model}\label{sec:MOR}
To obtain a reduced order model of system frequency dynamics we leverage model order reduction techniques to obtain a low-order dynamic model later used in predictive optimization of UFLS setpoints.
Several model order reduction techniques can be used to obtain a reduced order power system dynamics model. Some common techniques are balanced truncation, Krylov subspace method, and slow coherency-based aggregation \cite{Joe_Chow_MOR_book,Slow_Coherency}. In this work, we borrow techniques used in slow coherency-based aggregation to obtain an aggregated model similar to the SFR model but while capturing AC effects on frequency dynamics (SAFR model)\cite{Joe_Chow_MOR_book,Slow_Coherency,Dorfler_slow_coherency}. Slow coherency aggregation takes advantage of the phenomenon that generators that are more electrically coupled (coherent generators) tend to swing together exhibiting similar frequency oscillations \cite{Joe_Chow_MOR_book}. A coherent group exhibit fast intra-area oscillations within the coherent group and slower inter-area oscillations between the coherent groups. Therefore, a reduced-order model can be obtained by identifying coherent frequencies and representing a coherent group with its slow inter-area dynamics while neglecting the faster intra-area dynamics.

Similar to how the SFR model aggregates the full system dynamics into an equivalent generator, our proposed SAFR model aggregates the system into a single coherent group capturing the slow modes of the system while neglecting the faster oscillatory modes of the frequency dynamics. Note that the SAFR model obtained still captures AC effects and frequency dynamics, unlike the SFR model.
% \begin{figure}[] 
%     \centering\includegraphics[width=0.45\textwidth]{Figures/coherent_groups_2.jpg}
%     \caption{Frequency responses of two coherent generator groups observed in the 1648-bus test case.}
%     \label{fig:Coherent_groups}
%     \end{figure}

% \subsection{Simplified AC-aware Frequency Response Model (SAFR)}  
Given a dynamic model described by~\eqref{eqn:DAE_compact}, the system is first linearized around the power flow equilibrium. 
The linearized model is given by:
\begin{subequations}\label{eqn:Linearized DAE}
    \begin{align}
   \Delta \dot{x} &= \left. \frac{\partial f(x,y)}{\partial x} \right|_{x_0, y_0} \Delta x+ \left. \frac{\partial f(x,y)}{\partial y} \right|_{x_0, y_0}\Delta y \label{eqn:Linearized ODE}\\
        0 &= \left. \frac{\partial g(x,y)}{\partial x} \right|_{x_0, y_0} \Delta x + \left. \frac{\partial g(x,y)}{\partial y} \right|_{x_0, y_0} \Delta y, \label{eqn:Linearized Algebraic}
    \end{align}
\end{subequations}

\noindent where $\Delta x := x - x_0$ and $\Delta y := y - y_0$ are the deviations from the power flow equilibrium $x_0,y_0$\footnote{To simplify analysis, the left hand side of~\eqref{eqn:Linearized Algebraic} is kept as zero indicating no contingencies/load shedding added later in Section~\ref{sec:UFLS Setpoint Optimization}.}. The linearized model can be expressed compactly as
\begin{subequations}\label{eqn:Linearized DAE Compact}
    \begin{align}
        \Delta\dot{x} &= A_x\Delta x+ A_y\Delta y \label{eqn:Linearized ODE Compact} \\
        0 &= K_x\Delta x + K_y \Delta y. \label{eqn: Linearized Algebraic Compact}
    \end{align}
\end{subequations}

The angle and frequency states ($\Delta \delta, \Delta \omega$) of the linearized system in~\eqref{eqn:Linearized DAE Compact} are transformed into aggregated variables ($\Delta \delta_\text{r}, \Delta \omega_\text{r}$) and difference variables by defining a state transformation based on the model-order reduction procedure presented in~\cite{Joe_Chow_MOR_book,Slow_Coherency,Dorfler_slow_coherency}. The aggregated and difference variables capture the slow and fast modes of frequency oscillation respectively. The transformation matrix and its inverse are used to transform the system dynamics in terms of the aggregated and difference variables. Later on the difference variables are omitted to obtain a reduced order model that captures the slow aggregated system dynamics. 
 The aggregated variable $\Delta \delta_\text{r}$ is obtained by taking the inertia weighted average of the rotor angles,
\begin{align}\label{eqn:aggregated delta}
    \Delta \delta_{\text{r}} := \frac{\sum_{j\in \mathcal{G}} m_j\Delta\delta_{j}}{\sum_{j\in \mathcal{G}} m_j},
\end{align}
where $\Delta \delta_\text{r}$ is the aggregated variable that characterizes the overall trend in system frequency (i.e., the slow dynamics). 

Next, we define $z\in \mathbb{R}^{|\mathcal{G}|-1}$ as the difference variable that captures the faster modes of oscillation:   
\begin{align}\label{eqn:fast dynamics}
z_{i} := \Delta\delta_{i} -\Delta \delta_{k} \quad\forall i\in \mathcal{G} \backslash \{k\},
\end{align}
where $z_i$ is the relative rotor angle to a reference machine ($\Delta \delta_{k}$) for any arbitrary $k \in \mathcal{G}$. That is, $z$ captures the faster dynamics.
%The relative rotor angle is found by taking the difference of the rotor angles and a reference rotor angle. As a result, the overall trend shared by both terms (slow dynamics) is omitted leaving only the faster modes of oscillations (fast dynamics). %$z_i$ can be expressed as,

Together,~\eqref{eqn:aggregated delta} and ~\eqref{eqn:fast dynamics} define a state space transformation  of the system states ($\Delta \delta$ and $\Delta \omega$) in \eqref{eqn:Linearized DAE Compact} to the aggregated and difference variables, $\Delta \delta_\text{r}$ and $z$. 
The transformation shown  can be expressed in vector form as, 
\begin{align}\label{eqn:MOR state transformation}
\begin{bmatrix}
    \Delta\delta_\text{r}\\
    z
\end{bmatrix} = 
\begin{bmatrix}
    C\\
    G
\end{bmatrix}
\Delta \delta,
\end{align}
where $C\in \mathbb{R}^{1\times |\mathcal{G}|}$ and $G\in \mathbb{R}^{|\mathcal{G}|\times |\mathcal{G}|}$ are projection matrices that capture the transformations in ~\eqref{eqn:aggregated delta} and ~\eqref{eqn:fast dynamics}. In particular, $C := M_\text{a}^{-1}\mathbf{1}^{T}M$,
where $M\in \mathbb{R}^{|\mathcal{G}|\times |\mathcal{G}|}$ is a diagonal matrix with the machine inertia constants and $M_\text{a}\in \mathbb{R}$ is the aggregated system inertia and the transformation matrix $G$ is given by
\begin{align}
    G := \begin{bmatrix}
-1 & 1 & 0 & \cdots & 0 \\
-1 & 0 & 1 & \cdots & 0 \\
\vdots & \vdots & \vdots & \ddots & \vdots \\
-1 & 0 & 0 & \cdots & 1
\end{bmatrix}.
\end{align}
%Similarly, we can transform 
%$\Delta \omega:=\Delta \dot{\delta}$ to 
Taking the derivative of~\eqref{eqn:MOR state transformation} yields %:  $\Delta \omega_\text{r}:=\Delta \dot{\delta}_\text{r}$ and $\dot{z}$ ,
\begin{align}\label{eqn:MOR omega state transformation}
\begin{bmatrix}
    \Delta \omega_\text{r}\\
    \dot{z}
\end{bmatrix} =
\begin{bmatrix}
    C\\
    G
\end{bmatrix}
\Delta {\omega}.
\end{align}

% Similarly, the same transformation can be used to transform 
% We can denote the transformation in \eqref{eqn:aggregated delta} in vector form as follows:
% \begin{align}
%     \delta_\text{r} = T\Delta\delta_{\text{g}}:=M_\text{a}^{-1}U^TM\Delta\delta_{\text{g}}
% \end{align}
% where $M_\text{a}\in \mathbb{R}^{L\times L}$ is a diagonal matrix with the aggregated inertia of each group,$M\in \mathbb{R}^{N_\text{g}\times LN_\text{g}}$ is a diagonal matrix with the inertia coefficients of each generator in $\mathcal{G}$, and $U\in \mathbb{R}^{N_\text{g}\times L}$ is a grouping matrix whose elements satisfy the following:
% \begin{align}
%     U_{ij}=
%     \begin{cases}
%         1 & \text{if generator } i \text{ belongs to group } j\\
%         0 & \text{otherwise. } 
%     \end{cases}
% \end{align}
% % The aggregate variable $\delta_\text{r}$ is then used to represent the slow dynamics of the system.
% Similarly, for the fast dynamics within a coherent group, consider the rotor angle deviation from a reference angle within a coherent group,

% \begin{align}
%     \delta_{\text{r,}i}^f = \Delta\delta_{\text{g,}i} - \Delta\delta_{\text{g,}1} \forall i=2,,
% \end{align}

% where $\delta_{\text{r,}i}^f$ 
Next, the generator's turbine governor contributions are aggregated to obtain the aggregate governor contribution $\Delta P_\text{mr}\in \mathbb{R}$, and is given as
\begin{align}\label{eqn:MOR Governor aggregation}
    \Delta P_{\text{mr}} = \sum_{j\in \mathcal{G}}  \Delta P_{\text{m},j}.
\end{align}
The equivalent aggregated governor transfer function can be found by summing the individual governor transfer functions for all generators. In this work, we assume that the governor time constants $T_i$ are equal, simplifying the aggregation\footnote{If governor time constants are not identical, then we can still approximate the aggregated governor transfer functions as in~\cite{Joe_Chow_MOR_book}.}.
Applying the transformations shown in~\eqref{eqn:MOR state transformation}-\eqref{eqn:MOR Governor aggregation} to the system in~\eqref{eqn:Linearized DAE Compact} and omitting the difference variables $z$ results in a reduced-order system of the form
% \begin{subequations}
% \begin{align}
%     \Delta \dot{x}_\text{r} &= A_\text{r}\Delta x_\text{r}\\
%     \omega_\text{r} &= \left[0\quad I\quad 0 \right]x_\text{r}.
% \end{align}
% \end{subequations}
\begin{subequations}\label{eqn:Reduced Dynamics}
\begin{align}
    \Delta \dot{x}_\text{r} &= A_\text{r,x}\Delta x_\text{r} + A_\text{r,y}\Delta y \label{eqn:Reduced ODEs} \\
    0 &=K_\text{r,x} \Delta x_\text{r} + K_\text{y} \Delta y ,\label{eqn:Reduced Algebraic}
\end{align}
\end{subequations}
where $\Delta x_\text{r} := [\Delta \delta_\text{r} \quad \Delta \omega_\text{r} \quad \Delta P_\text{mr}]^T \in \mathbb{R}^{3}$ is the reduced dynamic state vector of the SAFR model. To include the effect of disturbances and UFLS action, consider the net change in real and reactive power injections : 
\begin{align}
    \Delta u[k] := \begin{bmatrix}
        P_\text{UFLS}[k] + \Delta P[k] \\ Q_\text{UFLS}[k] + \Delta Q[k],
        \end{bmatrix} \in \mathbb{R}^{2N}
\end{align}
where  $\Delta P[k](\Delta Q[k])$ is the active (reactive) power imbalance caused by a disturbance and $P_\text{UFLS}[k](Q_\text{UFLS}[k])$ and is active (reactive) load shedding amounts. The values of $P_\text{UFLS}[k] (Q_\text{UFLS}[k])$ are responsive to the frequency. The reduced order dynamics in~\eqref{eqn:Reduced Dynamics} can be re-written as follows:
\begin{subequations}\label{eqn:Reduced Dynamics with UFLS}
    \begin{align}
        \Delta\dot{x}_\text{r} &= A_\text{r,x}\Delta x_\text{r} + A_\text{r,y}\Delta y\label{eqn:Reduced ODE with UFLS}  \\
       \Delta u &= K_\text{r,x} \Delta x_\text{r} + K_\text{y} \Delta y\label{eqn:Reduced Algebraic with UFLS},
    \end{align}  
\end{subequations}
\noindent to include the change in power injections caused by UFLS action and the disturbance.

Since~\eqref{eqn:Reduced Algebraic with UFLS} is a linear system of equations and $K_\text{y}$ is non-singular, the algebraic variable $\Delta y$ can be eliminated by solving for $\Delta y$ and substituting it back into \eqref{eqn:Reduced ODE with UFLS} to obtain the following reduced-order dynamics:
%\begin{subequations} 
\begin{align}\label{eqn:Reduced Dynamics 2}   
    \Delta \dot{x}_\text{r} &= \left[ A_\text{r,x} - A_\text{r,y}K_\text{y}^{-1}K_\text{r,x}\right] \Delta x_\text{r} + A_\text{r,y}K_\text{y}^{-1}\Delta u,
\end{align}
%\end{subequations}
\noindent where $\Delta \omega_\text{r} = \left[0\quad 1\quad 0\right]\Delta x_\text{r}$.

% \noindent where $\Delta \omega_\text{r} = \left[\mathbf{0}_{ L \times L}\quad \mathbf{I}_{ L \times L}\quad \mathbf{0}_{ L \times L} \right]\Delta x_\text{r}$.

To optimize UFLS setpoints we consider a discretized version of  \eqref{eqn:Reduced Dynamics 2} by using trapezoidal integration with a timestep of $\Delta t$. The discrete timestep $k$ is related to continuous time $t$ as $ k = \floor{\frac{t}{\Delta t}}$. The discrete-time dynamics are then used in the optimization formulation to find the optimal UFLS setpoints.

% Similarly, the transformations in~\eqref{eqn:MOR state transformation}-\eqref{eqn:MOR Governor aggregation} can be applied to the system in~\eqref{eqn:Linearized DAE Compact} to obtain a system of the form,

% % With that, the linearized system equations in~\eqref{eqn:A_Matrix} can be reduced to the following:
% \begin{subequations}\label{eqn:Reduced Dynamics}
% \begin{align}
%     \dot{x}_\text{r} &= A_\text{r,x}\Delta x_\text{r} + A_\text{r,y}\Delta y \label{eqn:Reduced ODEs} \\
%     0 &=K_\text{x,r} \Delta x_\text{r} + K_\text{y} \Delta y \label{eqn:Reduced Algebraic}
% \end{align}
% \end{subequations}

% \noindent where $x_\text{r} = [\delta_\text{r} \quad \omega_\text{r} \quad P_\text{mr}]^T \in \mathbb{R}^{3M}$ is the reduced dynamic state vector and $y:=\left[\delta_{\text{b}}, V_{\text{b}}\right]^{T}$ represent the algebraic variables and is unreduced. Note that algebraic variables can be eliminated by solving for $y$ in~\eqref{eqn:Reduced Algebraic} and substituting in~\eqref{eqn:Reduced ODEs}, to obtain an equivalent reduced system model with the network dependencies embedded into the reduced dynamics and is shown below,
%  \begin{subequations}
%  \begin{align}
%     \dot{x}_\text{r} &= \left[ A_\text{r} - E_\text{r}K_\text{y}^{-1}K_\text{x,r}\right] x_\text{r} \\
%     \omega_\text{r} &= \left[0\quad I\quad 0 \right]x_\text{r}.
% \end{align}
% \end{subequations}
 
\begin{figure}[] 
    \centering
\includegraphics[width=\columnwidth]{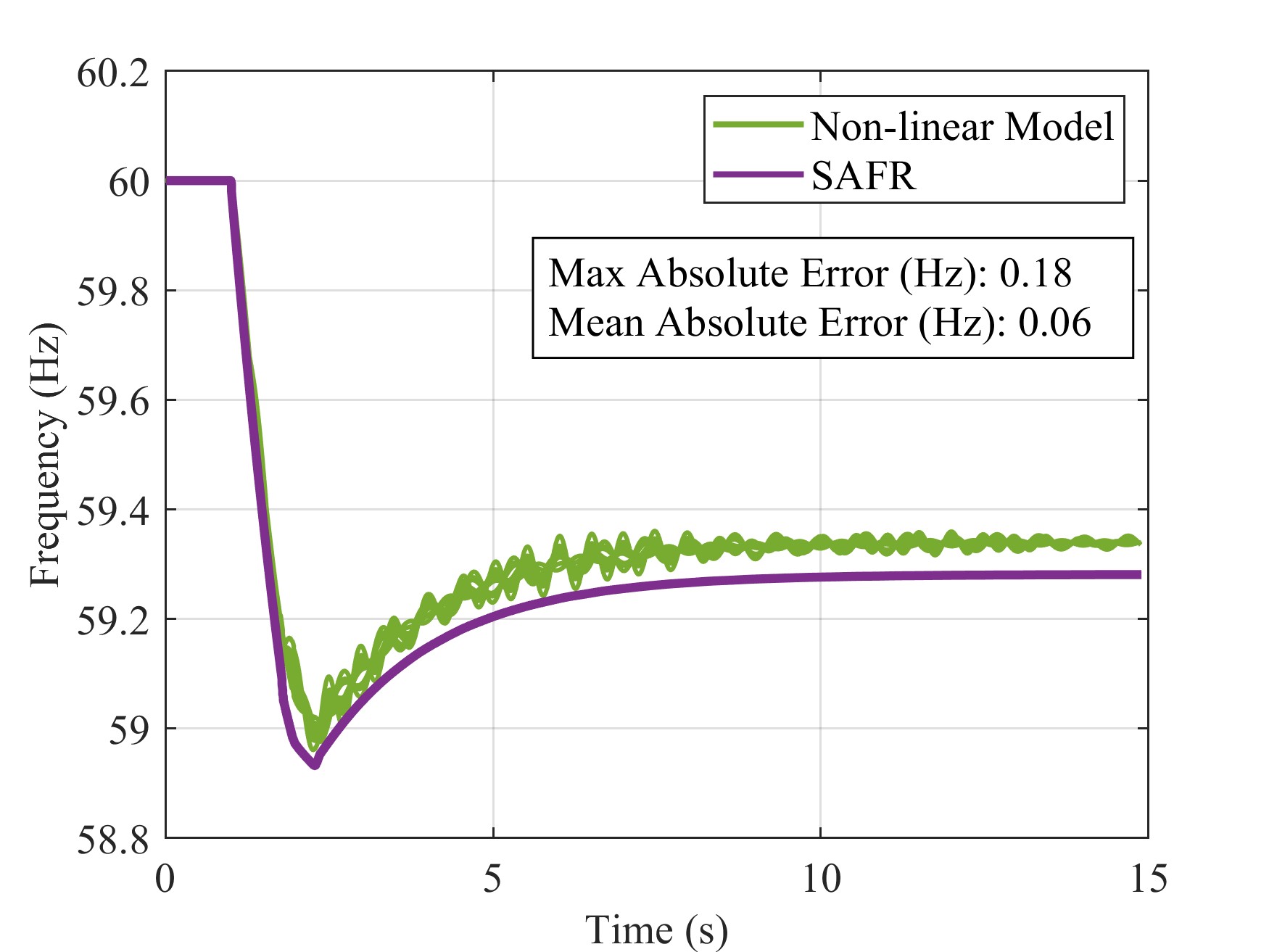}
    \caption{Comparison of frequency responses between the non-linear model (implemented in PSS/E) and the SAFR model.}
    \label{fig:Reduced Order Model}
    \end{figure}

The accuracy of the SAFR model is assessed by comparing the frequency response of the reduced order model to the full order non-linear model~\eqref{eqn:DAE_compact}. Fig~\ref{fig:Reduced Order Model} shows the frequency response of the full order model and the SAFR for 23 bus test-case provided by PSS/E for the same contingency and UFLS settings that were used in Fig~\ref{fig:COI Comparison}. The SAFR model  captures the frequency response of the non-linear full order model much closer with smaller error compared to the SFR model.

Next, we embed the SAFR model within a mixed-integer optimization formulation to predict the system's response to a disturbance and to determine optimal UFLS setpoints.
\section{UFLS Setpoint Optimization}\label{sec:UFLS Setpoint Optimization}
The SAFR model presented in Section \ref{sec:MOR} is used to predict the system's frequency response to a disturbance scenario with an imbalance of 25\%, as defined by NERC \cite{NERC}. The UFLS setpoints are then optimized such that to minimize the amount of load shed while ensuring the system frequency remains within acceptable limits. UFLS setpoints are defined by the load shed amounts at each load bus (usually expressed as a fraction of the present load) and the corresponding frequency threshold at which the load shed should occur ($\omega_\text{shed}$). 

\begin{figure}[]
    \centering\includegraphics[width=\columnwidth]{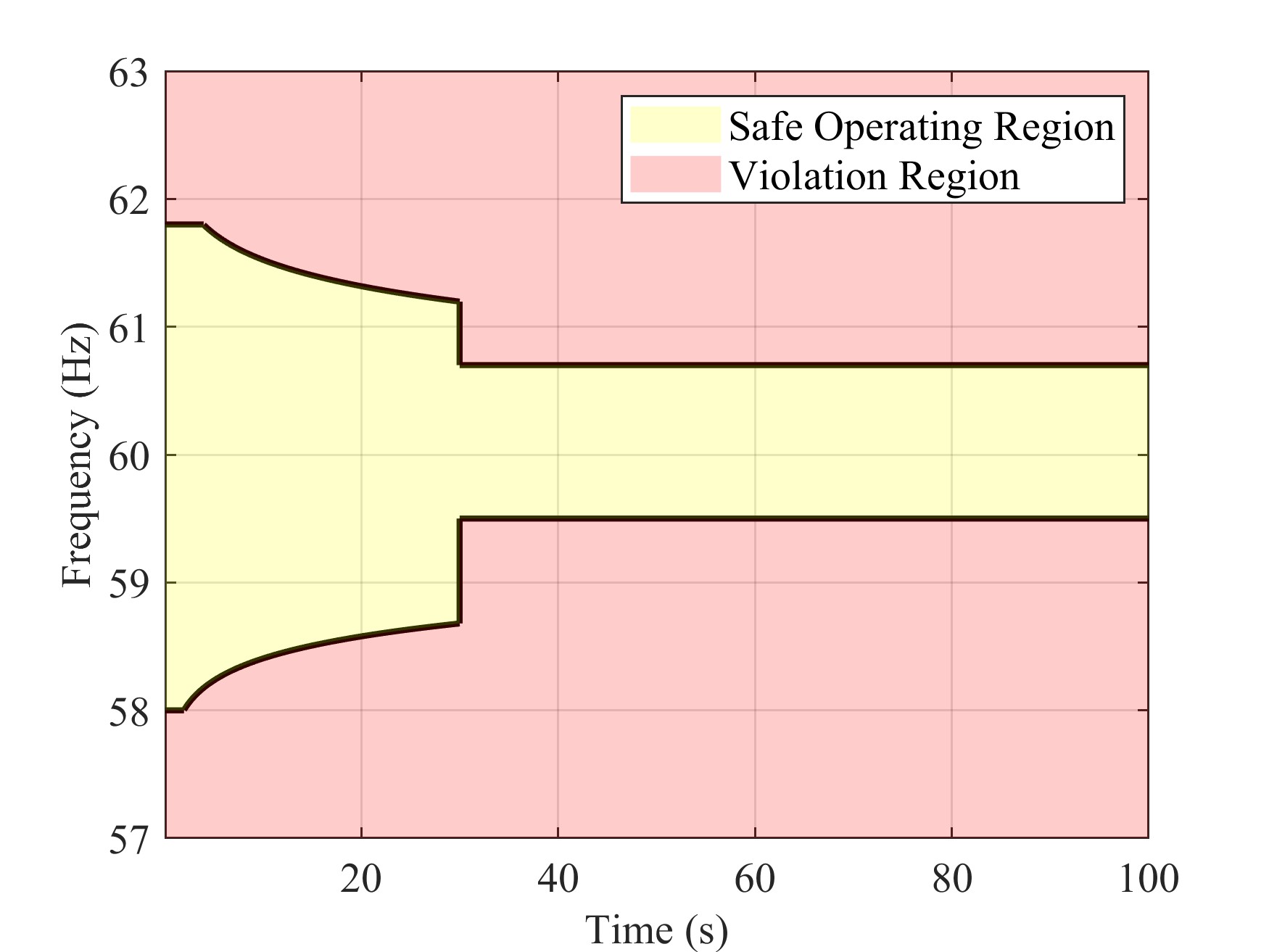}
    \caption{Frequency response design requirements for UFLS as outlined by NERC PRC-006-2.}
    \label{fig:NERC_frequency_band}
\end{figure}

\begin{figure}[]
    \centering
    \includegraphics[width=\columnwidth]{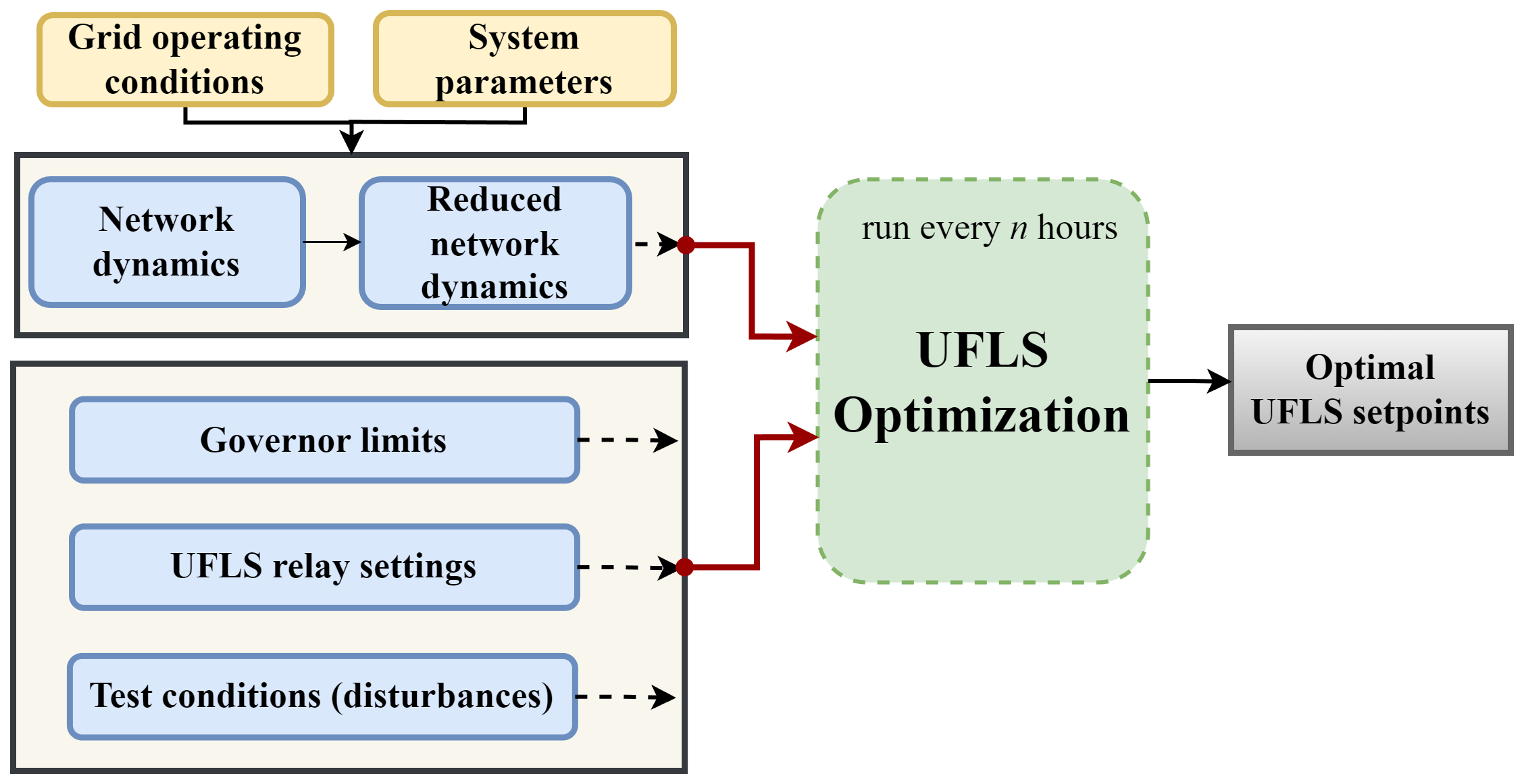}
    \caption{Block diagram of the UFLS optimization scheme.}
    \label{fig:methodology}
\end{figure}

Typically, a UFLS scheme is designed with multiple stages of load shedding with distinct frequency thresholds ($\omega_\text{shed}^i$) and load shedding amounts ($P_\text{shed}^i$) at each stage $i = 1, \hdots, N_\text{UFLS}$.
NERC defines a family of  Protection and Control (PRC) regulatory standards of which PRC-006-2 is the standard for UFLS~\cite{NERC}. PRC-006-2 defines the upper and lower bounds of what is considered to be 
safe frequency region as illustrated in Fig~\ref{fig:NERC_frequency_band}. A UFLS scheme, as specified by NERC, needs to ensure frequency remains within the safe region following a disturbance with an imbalance of up to 25\%  between generation and load. 
Furthermore, PRC-006-2 also defines the relay operating settings that define the 'time deadband' (200ms) which is how long the frequency should be below the threshold before UFLS is triggered. The time delay between UFLS trigger and actual relay tripping is also defined (100ms).  
Our proposed UFLS optimization scheme as illustrated in Fig~\ref{fig:methodology}, uses the presented SAFR model along with network information to adapt UFLS setpoints to changing grid conditions. Our proposed methodology aims to:
\begin{itemize}
    \item Minimize the amount of load shed while ensuring safe frequency recovery following a maximum credible contingency (25\% imbalance) and while considering the specified relay settings.
  
    \item Ensure no excessive load shedding at any given stage to avoid overshedding in the case of disturbances with smaller imbalances.

      \item Periodically update the UFLS setpoints based on the current system operating conditions.
      % {\color{blue}as illustrated in Fig~\ref{fig:methodology}.}
\end{itemize}
Next, we present the constraints associated with load shedding and turbine governors to optimize UFLS setpoints. 

\subsection{Load Shed Constraints}
For a UFLS scheme with $N_\text{UFLS}$ load shedding stages, the change in power injections at timestep $k$ caused by load shed corresponding to UFLS stage $i$ is given by $P_i^{\text{sh}}[k],Q_i^{\text{sh}}[k] \in \mathbb{R}^{N}$ respectively. Let the fraction of load shed at each load bus in UFLS stage $i$ and time step $k$ be $g_i[k]\in \mathbb{R}^{N}$ with the change in power real and reactive injections $P_i^{\text{sh}}[k]$ and $Q_i^{\text{sh}}[k]$ are definedby,
\begin{subequations}\label{eqn:Load Shedding Bi-linear}
\begin{align}
    P_i^{\text{sh}}[k] &:= g_i[k]P_{\text{d}}[k] \label{eqn:Load Shedding P}\\
    Q_i^{\text{sh}}[k] &:= g_i[k]Q_{\text{d}}[k] \label{eqn:Load Shedding Q},
\end{align}
\end{subequations}
\noindent where $P_{\text{d}}[k]$ and $Q_{\text{d}}[k]$ are defined in~\eqref{eqn:ZIP_load}. The load shedding fractions,$g_i[k]$, are used to determine the fraction of load to be shed at each load bus. Due to the voltage dependency of $P_{\text{d}}[k]$ and $Q_{\text{d}}[k]$~\eqref{eqn:Load Shedding Bi-linear} are non-convex. However, since bus voltages are bounded we can envelope $P_i^{\text{sh}}[k]$ and $Q_i^{\text{sh}}[k]$ as follows:
\begin{subequations} \label{eqn:Load Shedding Bounds}   
\begin{align}
    P_{i}^+[k] &:=  g_i[k]\left( P_{\text{P}} + \overline{V}I_{\text{P}} + \overline{V}^2Y_{\text{P}} \right) \\
    Q_{i}^+[k] &:= g_i[k]\left( Q_{\text{Q}} +\overline{V}I_{\text{Q}} +  \overline{V}^2Y_{\text{Q}} \right)\\
    P_{i}^-[k] &:= g_i[k]\left( P_{\text{P}} + \underline{V}I_{\text{P}} + \underline{V}^2Y_{\text{P}} \right)\\
    Q_{i}^-[k] &:= g_i[k]\left( Q_{\text{Q}} + \underline{V}I_{\text{Q}} + \underline{V}^2Y_{\text{Q}} \right),
\end{align}
\end{subequations}
 where $P_{i}^+[k]$, $Q_{i}^+[k]$ and  $P_{i}^-[k]$, $Q_{i}^-[k]$ are the upper and lower bounds\footnote{
 Note, that~\eqref{eqn:Load Shedding Bounds} assumes positive, inductive ZIP loads for notational convenience.
 }
 on $P_i^{\text{sh}}[k]$, $Q_i^{\text{sh}}[k]$, respectively. Upper and lower bounds on the total UFLS load shed input to the system, $P_\text{UFLS}^{+}[k]$ and $P_\text{UFLS}^{-}[k]$, are constrained as follows:
\begin{subequations}
\begin{align}
    P_\text{UFLS}^{+}[k+d] &:= \sum_{i=1}^{N_\text{UFLS}} P_{i}^+[k]  \label{eqn:P UFLS Power Upper} \\
    P_\text{UFLS}^{-}[k+d] &:= \sum_{i=1}^{N_\text{UFLS}} P_{i}^-[k]  ,\label{eqn:P UFLS Power Lower}
 \end{align}
\end{subequations}

\noindent where $d$ is the time delay between the relay triggering and the actual load shedding as specified by NERC~\cite{NERC}. $P_\text{UFLS}^{+}[k]$ and $P_\text{UFLS}^{-}[k]$ are constrained in a similar manner.Furthermore, we define upper and lower bounds on $\Delta u[k]$ at timestep k as,
\begin{subequations}
    \begin{align}
        \Delta u^{+}[k] &:= \begin{bmatrix}
            P_\text{UFLS}^{+}[k] + \Delta P \\ Q_\text{UFLS}^{+}[k] + \Delta Q
            \end{bmatrix} \\
            \Delta u^{-}[k] &:= \begin{bmatrix}
                P_\text{UFLS}^{-}[k] + \Delta P \\ Q_\text{UFLS}^{-}[k] + \Delta Q
                \end{bmatrix}.
    \end{align}
\end{subequations}
Similar constraints are added for $Q_\text{UFLS}^{+}[k]$ and $Q_\text{UFLS}^{-}[k]$. Due to the linear nature of the system model, applying upper and lower bounds on the true system input $\Delta u$[k] results in $\Delta \omega_\text{r}^{+}$ and $\Delta \omega_\text{r}^{-}$ that envelope the true system dynamics and are given by,
\begin{subequations}    \label{eqn:Reduced Dynamics Plus}
 \begin{align}
    \Delta \dot{x}_\text{r}^{+} &= \left[ A_\text{r,x} - A_\text{r,y}K_\text{y}^{-1}K_\text{r,x}\right] \Delta x_\text{r}^{+} + A_\text{r,y}K_\text{y}^{-1}\Delta u^{+} \\
    \Delta \omega_\text{r}^{+} &= \left[0\quad I\quad 0 \right]\Delta x_\text{r}^{+}
\end{align}
\end{subequations}
\noindent and 
\begin{subequations}    \label{eqn:Reduced Dynamics Minus}
 \begin{align}
    \Delta \dot{x}_\text{r}^{-} &= \left[ A_\text{r,x} - A_\text{r,y}K_\text{y}^{-1}K_\text{r,x}\right] \Delta x_\text{r}^{-} + A_\text{r,y}K_\text{y}^{-1}\Delta u^{-} \\
    \Delta \omega_\text{r}^{-} &= \left[0\quad I\quad 0 \right]\Delta x_\text{r}^{-}.
\end{align}
\end{subequations}
\subsection{Frequency Threshold Constraints}
Each stage, $i$, of load shed can only be triggered if the frequency is below the corresponding threshold, $\Delta \omega_\text{shed}^i$. In other words, the fraction of load shed $g_i[k]$ should be constrained to be zero unless the frequency is below the threshold. A binary variable $\alpha_i[k]\in \mathbb{R}$ is introduced to indicate whether or not frequency has dipped below the threshold corresponding to $i$-th UFLS stage. The following constraints are added to the optimization problem:
\begin{subequations} \label{eqn:Threshold Constraints}  
\begin{align}
    \alpha_i[k] &\leq 1 + \Delta w^i_\text{shed} - \Delta \omega_\text{r}^{+}[k]\\
    \alpha_i[k] &\geq \Delta w^i_\text{shed} - \Delta \omega_\text{r}^{+}[k] -\sum_{n=1}^{k-1} \alpha_i[n] \\
    \sum_{n=1}^{K}& \alpha_i[n] \leq K_\text{db} \label{eqn:Sum_alpha_db}\\
    \alpha_i[k] &\in \{0,1\},
\end{align}
\end{subequations}
    \noindent where $K_\text{db}$ is the time deadband of the relay. The constraints in~\eqref{eqn:Threshold Constraints} ensure that $\alpha_i[k]$ is zero if the reduced frequency is above $\omega_\text{shed}^i$. Note that in~\eqref{eqn:Threshold Constraints}, $ \Delta \omega_\text{r}^{+}[k]$ is used rather than $ \Delta \omega_\text{r}^{-}[k]$, since $ \Delta \omega_\text{r}^{+}[k] \geq \Delta \omega_\text{r}[k] \quad \forall k $, and when $ \Delta \omega_\text{r}^{+}[k]$ falls below the threshold then $ \Delta \omega_\text{r}[k]$ is also below a threshold and UFLS trigger is permissible. 
    % However, if $ \Delta \omega_\text{r}^{-}[k]$ is below a frequency threshold, $ \Delta \omega_\text{r}[k]$  may not necessarily be below the threshold and; therefore, using $ \Delta \omega_\text{r}^{-}[k]$ might cause impermissible premature UFLS actuation,which would not otherwise actuate.

    Then, $\alpha_i[k]$ constrains the fraction of load shed,
    \begin{subequations} \label{eqn:UFLS Trigger}       
    \begin{align}
    g_i[k+1] - g_i[k] &\leq \alpha_i[k-z] \quad \forall z=0\hdots K_\text{db}-1 \\
    g_i[0] &= 0,
    \end{align}
\end{subequations}
    \noindent which ensures that $g_i[k]$ remains zero until the frequency isbelow the frequency threshold for $K_\text{db}$ consecutive time steps. Furthermore, from~\eqref{eqn:Sum_alpha_db}, $\alpha_i[k]$ cannot be one for more than $K_\text{db}$ consecutive timesteps meaning that each element in $g_i[k]$ can only change (trigger) once throughout the entire time horizon, capturing the desired action of a UFLS relay.

\subsection{Frequency Nadir and Settling Frequency Constraints}
The proposed UFLS scheme must satisfy the requirements illustrated in Fig~\ref{fig:NERC_frequency_band}. We specifically ensure that the frequency nadir is within 58 Hz which is the lowest acceptable frequency, and that the settling frequency is within 59.5 Hz and 60.7 Hz. Since $\Delta \omega_\text{r}^{+}$ and $\Delta \omega_\text{r}^{-}$ bound the reduced frequency $\Delta \omega_\text{r}$, to satisfy the aforementioned requirements on frequency nadir the following constraints are added on $\Delta \omega_\text{r}^{+}$ and $\Delta \omega_\text{r}^{-}$,
\begin{subequations}\label{eqn:Settling Frequency Constraints}
    \begin{align}
        &\Delta \omega_\text{r}^{+}[k] \le \Delta \omega_\text{min} \quad \forall k\\
     &\Delta \omega_\text{r}^{-}[k] \le \Delta \omega_\text{min} \quad \forall k,
    \end{align}
\end{subequations}
\noindent where $\Delta \omega_\text{min}$ is the minimum allowed frequency (58 Hz). Similarly, to ensure the settling frequency requirements, the following constraints are added,
\begin{subequations}\label{eqn:Nadir Frequency constraints}
    \begin{align}
        \Delta \omega^\text{ss}_\text{min}\le &\Delta \omega_\text{r}^{+}[K] \le \Delta \omega^\text{ss}_\text{max}\\
        \Delta \omega^\text{ss}_\text{min}\le &\Delta \omega_\text{r}^{-}[K] \le \Delta \omega^\text{ss}_\text{max},
    \end{align}    
\end{subequations}

\noindent where $K$ is the time horizon considered in the optimization, $\omega^\text{ss}_\text{max}$ and$\omega^\text{ss}_\text{min}$ are the upper and lower settling frequency limits (60.7 and 59.5 Hz) respectively. Note that a large enough time horizon needs to be considered to ensure frequency settles. In this work, a time horizon of $K=\floor{\frac{15}{\Delta t}}$ was consistently sufficient for our chosen system

    % \begin{align}
    %     P_{i}[k] &= g_i[k]P_{\text{P}} + g_i[k]V_{\text{b}[k]}I_{\text{P}} + g_i[k]V^2_{\text{b}[k]}Y_{\text{P}}\\
    %     Q_{i}[k] &= g_i[k]Q_{\text{Q}} + g_i[k]V_{\text{b}[k]}I_{\text{Q}} + g_i[k]V^2_{\text{b}[k]}Y_{\text{Q}}
    % \end{align}

\subsection{Governor Limits Constraints}
\begin{figure}
    \centering
    \includegraphics[width=0.65\columnwidth]{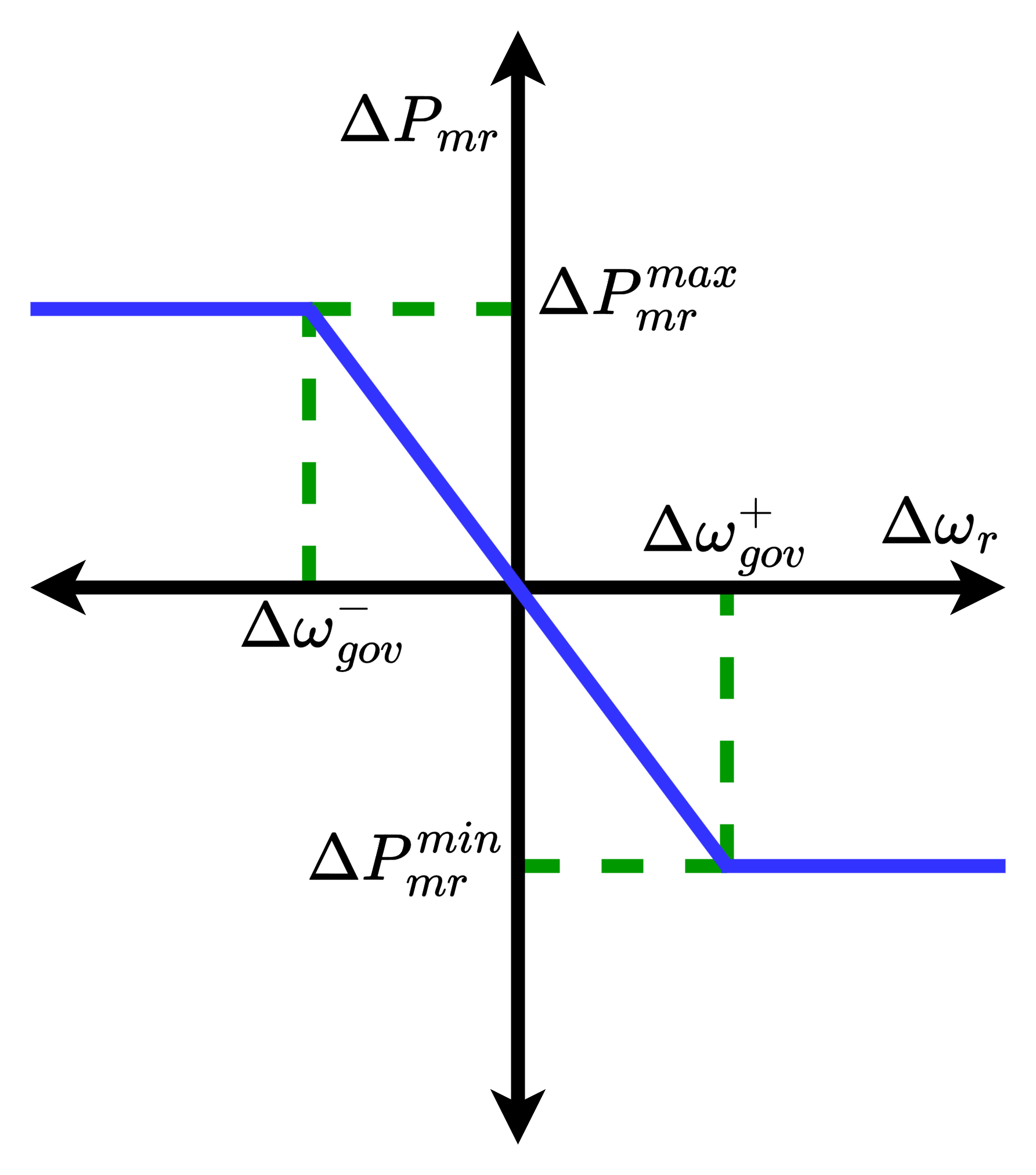}
    \caption{Illustration of Governor Limits.}
    \label{fig:Governor Limited}
\end{figure}

Turbine governors modulate a generator's mechanical power input based on its frequency.
Let the change in mechanical power for the systems shown in~\eqref{eqn:Reduced Dynamics Plus} and ~\eqref{eqn:Reduced Dynamics Minus} be denoted by  $\Delta P_\text{mr}^{+}[k]$ and $\Delta P_\text{mr}^{-}[k]$, respectively.
Turbine governors can only ramp up/down the mechanical power input within certain limits as illustrated in Fig~\ref{fig:Governor Limited}.  The frequency dynamics in~\eqref{eqn:Reduced Dynamics Plus} and~\eqref{eqn:Reduced Dynamics Minus} do not yet consider the governor saturation. Let $\Delta \Tilde{P}_\text{mr}^{+}[k]$, $\Delta \Tilde{P}_\text{mr}^{-}[k]\in\mathbb{R}$ denote the governor contribution after considering saturation:
\begin{subequations}  \label{eqn:Governor limits min max}  
\begin{align}
    \Delta \Tilde{P}_\text{mr}^{+}[k] &= \min \left\{\max \left\{\Delta {P}_\text{mr}^{+}[k],{\Delta  P}_\text{mr}^\text{min}\right\}, \Delta { P}_\text{mr}^\text{max}\right\}.
\end{align}
\end{subequations}
\noindent with similar constraints also being placed on $\Delta P_\text{mr}^{-}[k]$.
% \begin{subequations}  \label{eqn:Governor limits min max}  
% \begin{align}
%     P_\text{m}^{+}[k] &= \min \left\{\Tilde{P}_\text{m}^{+}[k], \overline{P_\text{m}}\right\} \land 
%     P_\text{m}^{-}[k] =  \min \left\{\Tilde{P}_\text{m}^{-}[k], \overline{P_\text{m}}\right\}\\
%     P_\text{m}^{+}[k] &= \max \left\{\Tilde{P}_\text{m}^{+}[k],\underline{P_\text{m}}\right\} \land 
%     P_\text{m}^{-}[k] =  \max \left\{\Tilde{P}_\text{m}^{-}[k],\underline{P_\text{m}}\right\},
% \end{align}
% \end{subequations}
% where $\overline{P_\text{m}}$ and $\underline{P_\text{m}}\in \mathbb{R}^M$ are the upper and lower bounds on the governor contributions respectively.

% The constraints in~\eqref{eqn:Governor limits min max} ensure that the governor contributions do not exceed their limits. 

To engender a linear formulation of ~\eqref{eqn:Governor limits min max} we introduce the binary variable $\gamma^{+}[k]\in\left\{0,1\right\}$ ~\eqref{eqn:Governor limits min max} constrained as follows:
 \begin{align}
     \frac{\Delta { P}_\text{mr}^\text{max}-{\Delta P}_\text{mr}^{+}[k]}{S} & \le \gamma^{+}[k] \le 1+\frac{\Delta { P}_\text{mr}^\text{max}-{\Delta P}_\text{mr}^{+}[k]}{S} \label{eqn:Governor Gamma Constriant Plus}
\end{align}

\noindent where the value of $S$ is chosen such that the expressions on the left and right hand side of~\eqref{eqn:Governor Gamma Constriant Plus} neither exceed one nor fall below zero, respectively. In our application,  $S=500$ was determined to be sufficiently large. $\gamma^{+}[k]$ is constrained by~\eqref{eqn:Governor Gamma Constriant Plus} to be zero when governor saturates and vice versa. Therefore,  $\gamma^{+}[k]$ is used to constrain $ \Delta \tilde{P}_\text{mr}^{+}[k]$ as,

\begin{align} \label{eqn:Governor limits binary plus}
 \Delta \tilde{P}_\text{mr}^{+}[k] = \gamma^{+}[k]&\Delta P_\text{mr}^{+}[k] + \left(1- \gamma^{+}[k]\right)\Delta { P}_\text{mr}^\text{max} .
 % \gamma^{+}[k] &\in \left\{0,1\right\} \quad \forall k.
\end{align}

% \begin{subequations} \label{eqn:Governor limits binary minus}    
%     \begin{align} 
%      \tilde{P}_\text{m}^{-}[k] &= \gamma^{-}[k]P_\text{m}^{-}[k] + \left(1- \gamma[k]\right)\overline{P_\text{m}} \\
%      \frac{\overline{P_\text{m}}-{P}_\text{m}^{-}[k]}{L} & \le \gamma^{-}[k] \le 1+\frac{\overline{P_\text{m}}-{P}_\text{m}^{-}[k]}{L} \label{eqn:Governor Gamma Constriant Minus} \\
%      \gamma^{-}[k] &\in \left\{0,1\right\} \quad \forall k.
%     \end{align}
%     \end{subequations}

% {\color{blue} The value of $S$ is chosen such that the fractions shown in~\eqref{eqn:Governor Gamma Constriant Plus} are within the range $[-1,1]$. A worst-case value of $\Delta P_\text{mr}^{+}[k]$ can be obtained from the governor droop gain and worst-case frequency deviation (56 Hz is used in this work). The value of $S$ is then chosen such that $S\gg\Delta P_\text{mr}^{+}[k]$ ensuring that the fractions in ~\eqref{eqn:Governor Gamma Constriant Plus} are within $[-1,1]$.} 
Similar constraints are introduced to capture the upper limits on $\Delta P_\text{mr}^{-}[k]$ using a binary variable $\gamma^{-}[k]$. The governor's lower limits can be captured similarly although for brevity, we do not include them here.
However, the constraints in~\eqref{eqn:Governor limits binary plus} do contain bi-linear terms resulting in increased computational complexity. Therefore, the bi-linear terms in~\eqref{eqn:Governor limits binary plus}  can be relaxed as follows:
\begin{subequations} \label{eqn:Governor limits relaxed}    
    \begin{align} 
    &\Delta  \tilde{P}_\text{mr}^{+}[k] = t^{+}[k] + \left(1- \gamma^{+}[k]\right)\Delta { P}_\text{mr}^\text{max} \\
    &-\gamma^{+}[k] S \le t^{+}[k] \le \gamma^{+}[k] S\\
    &t^{+}[k]\ge -(1-\gamma^{+}[k]) S + \Delta P_\text{mr}^{+}[k] \\ &t^{+}[k]\le (1-\gamma^{+}[k]) S + \Delta P_\text{mr}^{+}[k].
    \end{align}
    \end{subequations}
   A similar relaxation is done for $\Delta {P}_\text{mr}^{-}[k]$ as well.  Note that due to the binary nature $\gamma^{+}[k]$ and $\gamma^{-}[k]$  the relaxation in~\eqref{eqn:Governor limits relaxed} is exact. Hence, $\Delta \tilde{P}_\text{mr}^{+}[k]$ and $\Delta \tilde{P}_\text{mr}^{-}[k]$ accurately capture the governor outputs when the governor saturates. 

   \subsection{Overshedding Constraints}
   To ensure that the UFLS scheme does not shed excessive amounts of load at any given stage, the total fraction of load shed at each stage is constrained as follows:
    \begin{subequations} \label{eqn:Overshedding Constraints}
    \begin{align}
        {g_i[k]^T P_\text{d,0}} \le \overline{g}\left({1^TP_\text{d,0}} \right) \quad \forall i,
        \end{align}
    \end{subequations}
    \noindent where $P_\text{d,0}$ is the initial power demand, and $\overline{g}$ is the maximum fraction of load that can be shed at any given stage (7.5\% as specified by NERC). Furthermore, the frequency thresholds ($ \Delta\omega^i_\text{shed}$) are constrained as follows:
    \begin{subequations} \label{eqn:Frequency Threshold seperation}
    \begin{align}
        \Delta \omega_\text{shed}^i \leq \overline{\Delta \omega}_\text{shed}\\
       \Delta \omega_\text{shed}^{i+1} - \Delta \omega_\text{shed}^i \ge {\Delta \omega}_\text{sep},
        \end{align}
    \end{subequations}
    \noindent where ${\Delta \omega}_\text{sep}$ is the minimum separation between the frequency thresholds and $\overline{\Delta\omega}_\text{shed}$ the largest allowable frequency threshold. This is to ensure that, 1) frequency thresholds are not too close to the nominal frequency to avoid premature and maybe unnecessary load shedding, and 2) the frequency thresholds are spaced enough apart. Lighter disturbances that would have only needed a single stage of UFLS would trigger multiple stages of UFLS if the frequency thresholds are too close together. The values defined by NERC for ${\Delta \omega}_\text{sep}$ and $\overline{\Delta\omega}_\text{shed}$ (i.e. 0.2Hz and 59.5Hz) are used in this work.  
\subsection{Overall Formulation}
The overall UFLS optimization formulation can be summarized as follows: 
\begin{equation}
\label{eqn:P1}
\text{\textbf{P1:}} \quad
\begin{aligned}
    & \underset{\omega^{i}_\text{shed}, g_i[k]}{\text{minimize}}
    & & {g_i[k]^T P_\text{d,0}} \\[5pt]
    & \text{subject to}
    & & \eqref{eqn:Load Shedding Bounds} - \eqref{eqn:Governor Gamma Constriant Plus} \text{ and } \eqref{eqn:Governor limits relaxed} -  \eqref{eqn:Frequency Threshold seperation}    .
\end{aligned}
\end{equation}

The implementation is summarized in Algorithm \ref{Algm_1}. It begins by linearizing the non-linear system model and deriving the SAFR model. For every $p$ hours, the algorithm solves the optimization problem to calculate the UFLS setpoints, adapting to changing grid conditions.

\begin{algorithm}[]
\caption{{\bf UFLS Setpoint Optimization} \label{Algm_1}}
    \SetKwInOut{Input}{Input}
    \SetKwInOut{Output}{Output}
    \Input{Network parameters, governor limits, UFLS relay settings, disturbance scenario}
    \Output{Optimized UFLS setpoints}
    \textbf{Step 1:} Instantiate Models \\
    \Indp
    \textbf{Linearize} system dynamic model as shown in ~\eqref{eqn:Linearized DAE Compact} \\
    \textbf{Obtain} the SAFR model from~\eqref{eqn:Reduced Dynamics 2} \\
    \Indm
    \textbf{Step 2:} Time-domain Optimization \\
    \textbf{Initialize} $n = 1$\\
    \While{$\text{True}$}{   
            \textbf{Solve:} \textbf{P1} \\
            $n \gets n + 1$\\
            \textbf{Goto:} Step 1\\
        }
    % \textbf{Step 3}: Compile all UFLS setpoints\\
\end{algorithm}

\section{Simulation Results}\label{sec:Simulation_Res}
\begin{figure}[] 
    \centering
\includegraphics[width=0.45\textwidth]{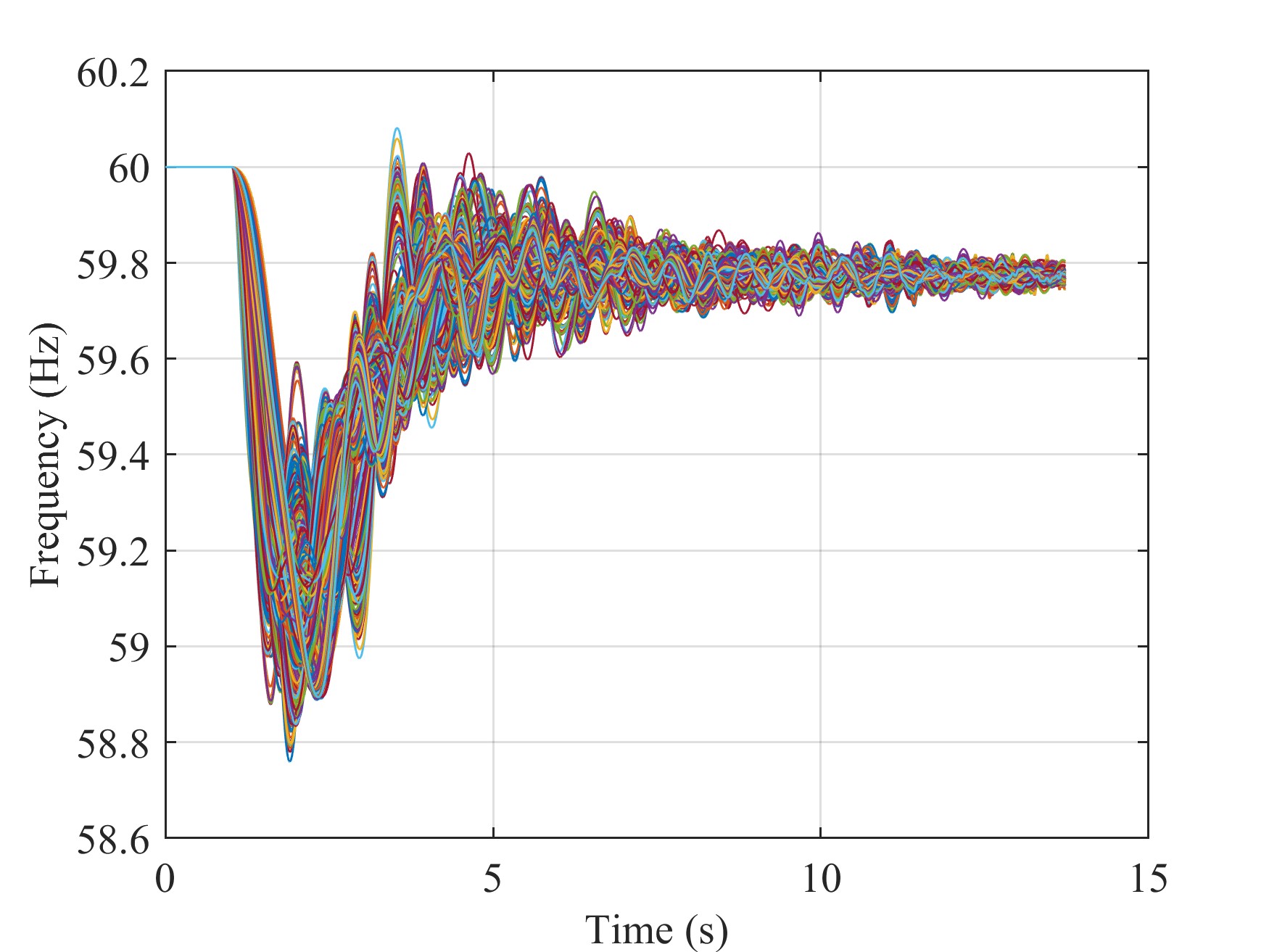}
    \caption{Frequency responses of all buses to 25\% imbalance with SAFR-based UFLS setpoints.}
    \label{Fig:full_res}
\end{figure}

The proposed UFLS scheme is validated on the 1648-bus test case provided by PSS/E \cite{Siemens2018}. The UFLS setpoints were optimized for a disturbance scenario with an imbalance of 25\% caused by generator tripping. A timestep ($\Delta t$) of 0.01s is considered, along with a governor droop gain ($R_i$) of 0.05 and a governor time constant ($T_i$) of 0.1s. The governor's limits were assigned assuming generators have 15\% reserves. The UFLS setpoints were then obtained by solving~\eqref{eqn:P1} and were then implemented on the full non-linear system using PSS/E. The optimization was conducted using MacBook Pro, M-1 Pro chip, with Gurobi as the MILP solver. The response of the system is shown in Fig~\ref{Fig:full_res}. The response shown in Fig~\ref{Fig:full_res}, and satisfies the performance criteria laid out by NERC.

However, with the ongoing evolution of grid conditions towards reduced inertia and increased DER penetration, it is important to evaluate and compare the robustness and adaptability of various UFLS schemes under diverse operating scenarios. To this end, the performance of the proposed SAFR-based UFLS scheme is assessed against a conventional static UFLS scheme and an SFR-based UFLS scheme. To determine conventional static UFLS setpoints, UFLS relays are added to various load buses until the load-shedding requirement specified by NERC is met~\cite{NERC}. The setpoints are then tested by simulating a 25\% imbalance disturbance and ensuring the frequency response satisfies specified requirements. Note that UFLS relays come with additional logic that stops relay tripping in case of net-generation behind the relay (net-gen blocking). This is to avoid a counter-productive loss of generation during an under-frequency event. Therefore, the implemented conventional UFLS scheme has the net-gen blocking logic embedded. A 1648-bus test case, provided by PSS/E, is considered. The test case consists of 1648 buses with 285 generators. The performance of the UFLS schemes are evaluated on the test case under three different scenarios: a base case, a reduced inertia case, and high DER penetration, reflecting the different power systems operating conditions.
\subsection{Base Case}
\begin{figure}[] 
    \centering
    \includegraphics[width=0.45\textwidth]{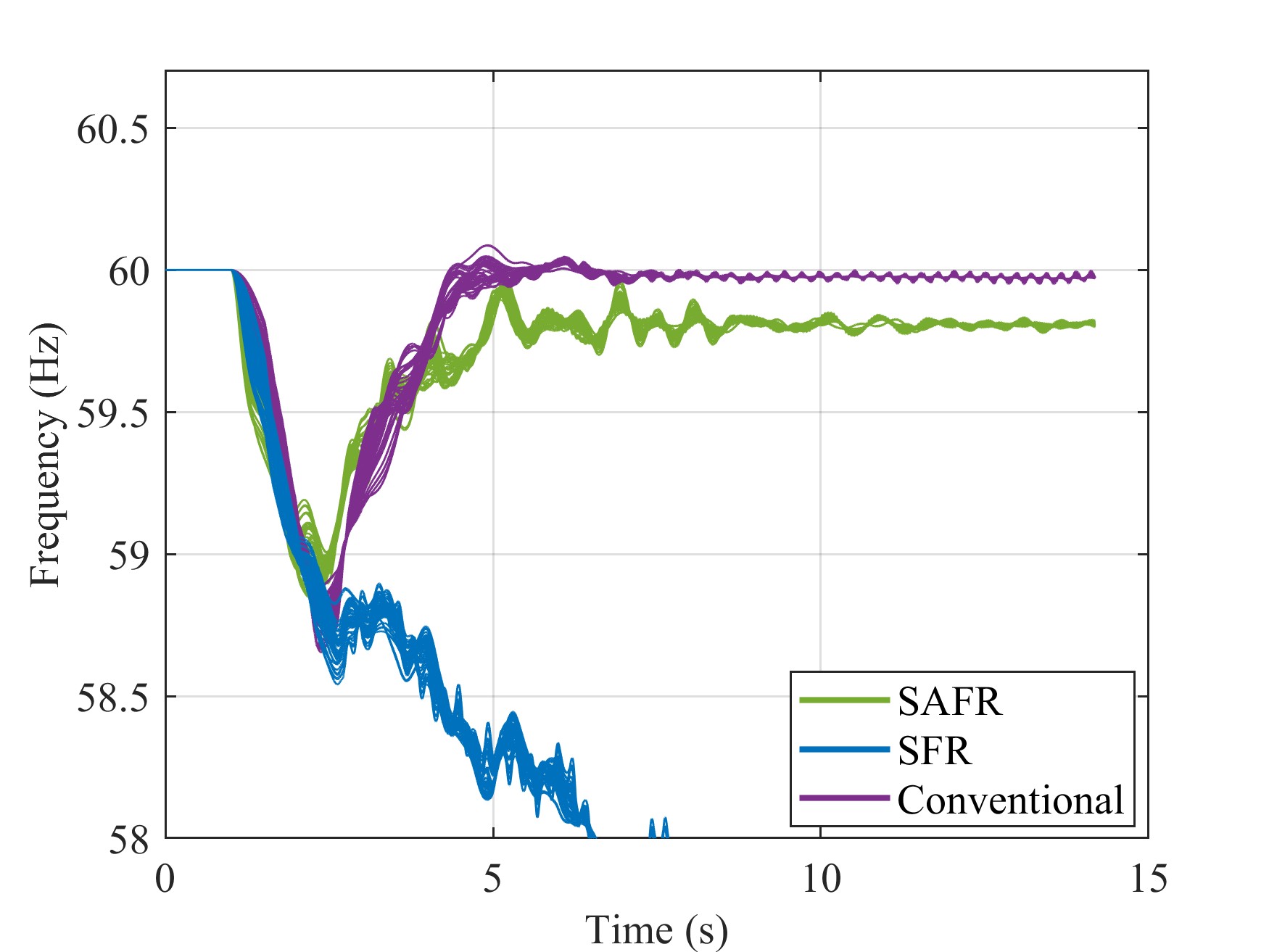}
    \caption{Comparison of simulation results for the base case scenario using SAFR, SFR, and the conventional method.}
    \label{Fig:100inertia_no_der}
\end{figure}

In this scenario, the network operates under normal loading conditions. The largest credible contingency with   25\% imbalance is introduced by simulating the loss of generators. This disturbance is applied at $t = 1$s to evaluate the performance of the proposed SAFR scheme in comparison to conventional and SFR-based UFLS schemes. 
The SAFR model captures AC-network effects and accounts for changes in load caused by voltage variations. The higher accuracy yields UFLS set points that reflect these effects, offering greater accuracy compared to the SFR model, which neglects voltage dependency. The results as illustrated in Fig. \ref{Fig:100inertia_no_der} demonstrate that the SFR-based UFLS approach under-estimates the required amount of load shed needed to halt frequency decline. Consequently, the method failed to stabilize the frequency under this scenario. In contrast, both the proposed SAFR and conventional approach shed appropriate amounts of load and stabilized frequency within safe limits. However, the settling frequency achieved by the conventional scheme was higher than that of the SAFR-based approach. This indicates that the conventional scheme sheds more load than necessary to stabilize the system as shown in Table \ref{tab:normal_case}. The solve times needed to obtain the SAFR and SFR-based UFLS setpoints are also shown in Table \ref{tab:normal_case}. The optimization formulation presented in~\eqref{eqn:P1} is larger than the SFR-based UFLS optimization due to the additional variables introduced to envelop the system dynamics and capture governor saturation. As such the solve time needed to obtain the UFLS setpoints using the SAFR model is larger. However, the solve times are still within 5 minutes for the 1649 bus test-case. Note, for clearer illustration, only a subset of the bus frequencies is plotted. 

\begin{table}[]
\caption{Comparison of UFLS Schemes: Base Case}
\begin{center}
\setlength\tabcolsep{10pt}
 \centering
 %\normalsize
 \label{tab:normal_case}
\begin{tabular}{lcccc}
\toprule
{\textbf{Model}} 
 & \textbf{Solver}& \textbf{Nadir}& \textbf{Settling} &\textbf{Load} \\

 {}& \textbf{Time}& \textbf{Frequency}& \textbf{Frequency} &\textbf{Shed} \\
  {}& \textbf{(s)}& \textbf{(Hz)}& \textbf{(Hz)} &\textbf{(\%)} \\
 \toprule
SAFR & 251.71& 58.75& 59.79& 16.79                    \\
SFR  & 0.15 & - & - & 15.58 \\
CONV &  - & 58.65 & 59.97 & 21.92 \\
  \bottomrule
\end{tabular}
\end{center}
\end{table}

\subsection{Reduced Inertia System}
To conduct a more comprehensive performance comparison of the different UFLS schemes, the system parameters were modified to represent a reduced inertia scenario. In this setup, the inertia of each synchronous generator is reduced to half of its original value. The performance of the UFLS schemes under the reduced inertia system is illustrated in Fig~\ref{Fig:100inertia_no_der}. Similar to the base case, the SFR-based UFLS approach failed to stabilize the system. Meanwhile, the conventional UFLS scheme stabilizes the system but at the cost of shedding approximately 10\% more load than the proposed SAFR-based UFLS scheme, as recorded in Table~\ref{tab:less_inertia}. Notably, the proposed SAFR scheme outperformed the conventional approach by maintaining the system frequency within the safe operating range while minimizing load shedding, thereby demonstrating its superior efficiency and adaptability.
\begin{figure}[]
    \centering
    \includegraphics[width=0.45\textwidth]{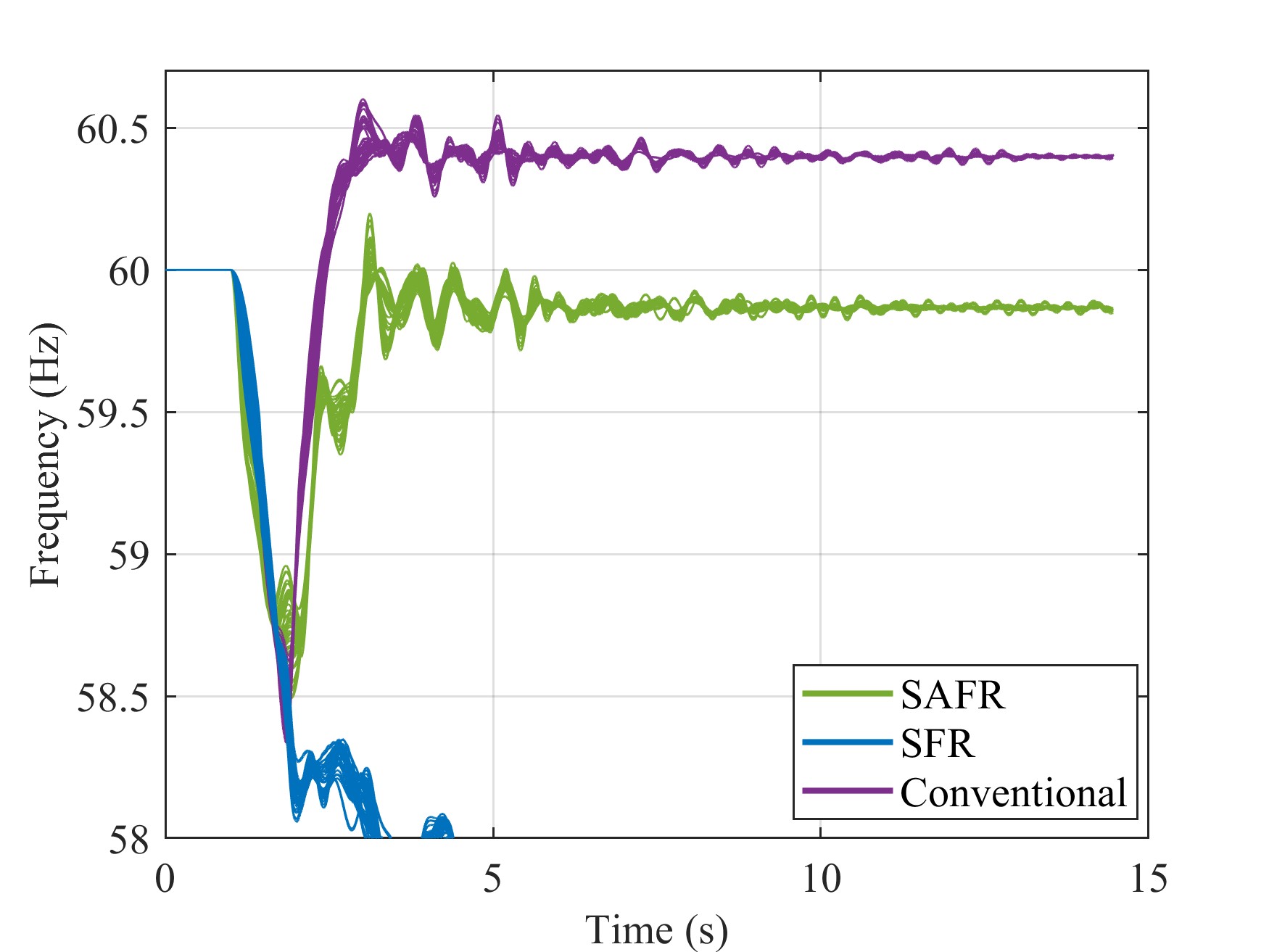}
    \caption{Comparison of simulation results for the reduced inertia scenario using SAFR, SFR, and the conventional method.}
    \label{fig:50inertia_noder}
\end{figure}

\begin{table}[htbp]
\caption{Comparison of UFLS Schemes: Reduced Inertia}
\begin{center}
\setlength\tabcolsep{10pt}
 \centering
 %\normalsize
 \label{tab:less_inertia}
\begin{tabular}{lcccc}
\toprule
{\textbf{Model}} 
 & \textbf{Solver}& \textbf{Nadir}& \textbf{Settling} &\textbf{Load} \\

 {}& \textbf{Time}& \textbf{Frequency}& \textbf{Frequency} &\textbf{Shed} \\
  {}& \textbf{(s)}& \textbf{(Hz)}& \textbf{(Hz)} &\textbf{(\%)} \\
 \toprule
SAFR & 64.92          & 58.49                 & 59.86                    & 17.47              \\
SFR  & 0.07 & - & -& 14.14   \\
CONV & - & 58.33                 & 60.39                    & 26.64 \\
  \bottomrule
\end{tabular}
\end{center}
\end{table}

\subsection{Penetration of DER}
\begin{figure}[]
    \centering
    \includegraphics[width=0.45\textwidth]{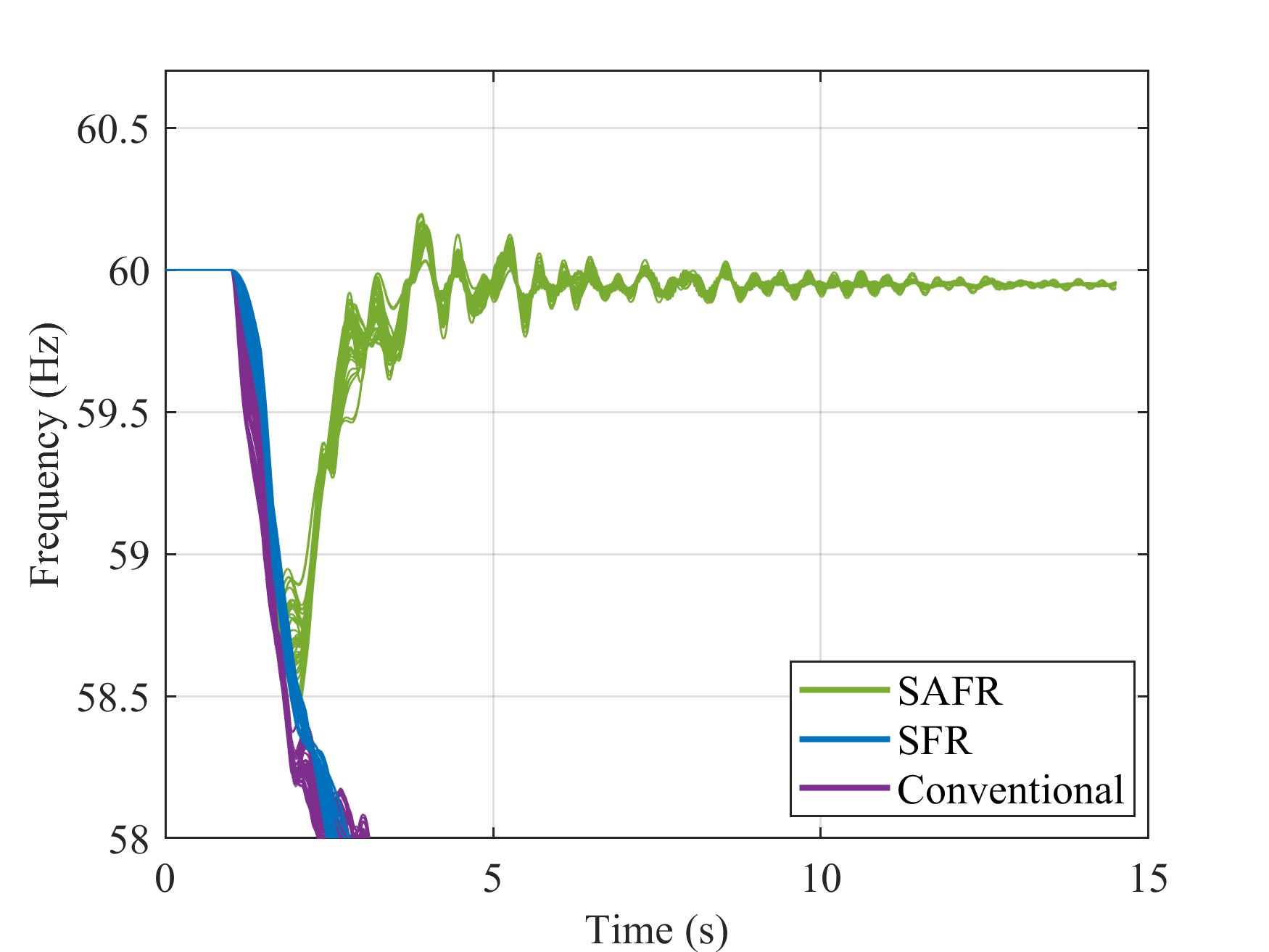}
    \caption{Simulation results for high DER penetration.}
    \label{fig:50inertia_50_DER}
\end{figure}

We modified the network further to evaluate the performance of UFLS schemes in systems with high DER penetration and reduced inertia. This modification involved incorporating DERs to account for 20\% of the total load while keeping the 50\% reduction in system inertia. The results for this scenario are presented in Fig~\ref{fig:50inertia_50_DER}. From Fig~\ref{fig:50inertia_50_DER}, it is evident that both the SFR-based and conventional UFLS schemes failed to shed the necessary amount of load required to stabilize the system under these conditions. However, the proposed UFLS setpoints from the SAFR-based UFLS scheme successfully stabilized the frequency of the system within safe limits.  Table~\ref{tab:penetration_der} summarizes the \% load shed, frequency nadir and settling frequency under the different UFLS schemes.

\begin{table}[htbp]
\caption{Comparison of UFLS Schemes: DER Penetration}
\begin{center}
\setlength\tabcolsep{10pt}
 \centering
 \label{tab:penetration_der}
\begin{tabular}{lcccc}
\toprule
{\textbf{Model}} 
 & \textbf{Solver}& \textbf{Nadir}& \textbf{Settling} &\textbf{Load} \\

 {}& \textbf{Time}& \textbf{Frequency}& \textbf{Frequency} &\textbf{Shed} \\
  {}& \textbf{(s)}& \textbf{(Hz)}& \textbf{(Hz)} &\textbf{(\%)} \\
 \toprule
SAFR & 13.29 & 58.46                 & 59.95                    & 18.96                    \\
SFR &  0.12 & -                      & -                       & 15.66   \\                
CONV  & - & -                      & -                       & 14.06   \\
  \bottomrule
\end{tabular}
\end{center}
\end{table}

\subsection{Multiple disturbance scenarios}
\begin{figure}[] 
    \centering
    \includegraphics[width=0.5\textwidth]{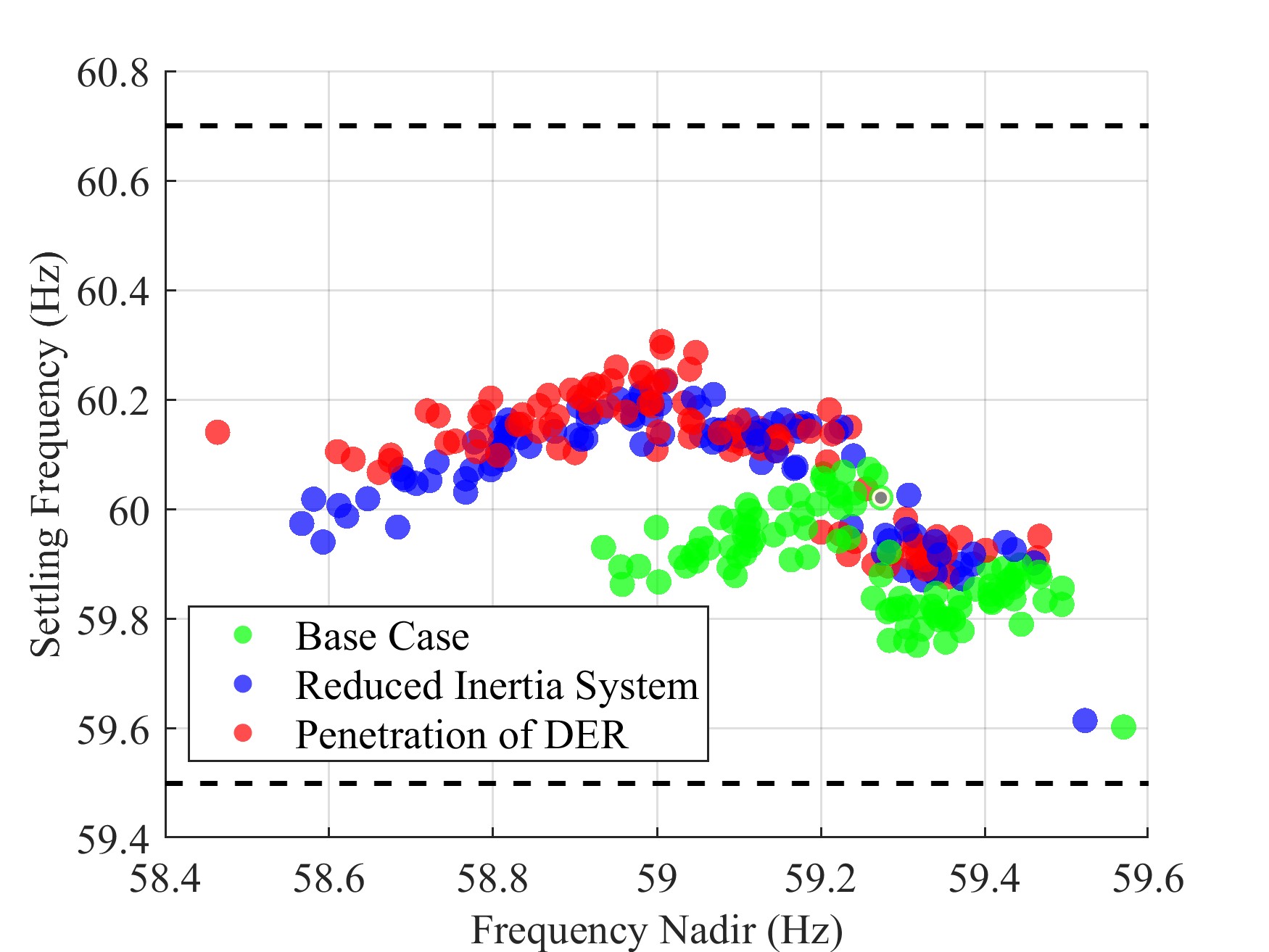}
    \caption{Simulation results for different scenarios.}
    \label{fig:mulitple_dist}
\end{figure}

To further demonstrate the efficacy of the proposed SAFR-based approach, simulations were conducted under various disturbance scenarios to highlight the robustness of the proposed methodology in handling diverse disturbances. These scenarios involve the application of various disturbance scenarios with imbalances ranging from 5\% to 25\%. For this analysis, 100 distinct disturbances were identified. For all 100 disturbance scenarios, the UFLS setpoints obtained from the proposed SAFR-based approach were implemented, and the resulting frequency nadir and settling frequency are illustrated in Fig.~\ref{fig:mulitple_dist}.
The results, presented in Fig. \ref{fig:mulitple_dist}, illustrate the system's performance under three distinct operating scenarios and various disturbance magnitudes. These results confirm that the proposed SAFR approach successfully stabilizes the system while adhering to NERC's operational guidelines.
\section{Conclusion and Future Work}\label{sec:Conclusion}
In this study, slow coherency aggregation techniques are leveraged to obtain an SAFR model of frequency dynamics used to optimize UFLS setpoints. Previous methods have relied on the SFR model to predict system frequency response; however, the SFR model is agnostic to the dependency of load and generator injections on voltage, which significantly affects system behavior during under-frequency events. Unlike the SFR model, the SAFR model is a predictive model that incorporates AC network effects such as voltage dependencies, providing a more accurate representation of the system's dynamic response. Furthermore, unlike prior work that neglected governor limits, we included these constraints in our modeling. UFLS applications involve scenarios with large contingencies that can cause governors to saturate, and failure to account for these limits can lead to substantial inaccuracies in predicting frequency response.

The performance of our SAFR-based UFLS setpoints under the largest credible contingency scenario was verified using PSS/E. The SAFR-based UFLS setpoints were also compared to setpoints derived from the SFR model and conventional static UFLS setpoints. Our results demonstrated that the proposed UFLS setpoints successfully restored system frequency to safe levels with minimal load shedding. In contrast, the SFR-based setpoints significantly underestimated the required load shedding, failing to halt the frequency decline. While the conventional setpoints performed adequately under the base case, they were ineffective in situations with reduced inertia and high penetration of DERs, where the static setpoints failed to arrest the frequency decline. 

Future work will incorporate higher-order generator models and exciters to further enhance the fidelity of the system representation. Furthermore, the effect of uncertainty in available load behind load buses would be incorporated to obtain UFLS setpoints that are robust against said uncertainty.

\bibliographystyle{ieeetr}
\bibliography{references.bib}

\end{document}